\documentclass[
aps,prb,twocolumn,superscriptaddress,showkeys,showpacs
]{revtex4-1}
\usepackage{amsmath,amsfonts,amssymb,amsthm,epsfig,epstopdf,url,array}
\usepackage{dsfont}
\usepackage{bm}
\usepackage{dcolumn}
\usepackage{mathrsfs}
\usepackage{latexsym}
\usepackage{graphics}
\usepackage{textcomp}
\usepackage{mathrsfs}
\usepackage{indentfirst}
\usepackage{fancyhdr}
\usepackage{titletoc}
\usepackage{titlesec}
\usepackage{float}
\usepackage{ragged2e}
\usepackage{verbatim}
\usepackage{multirow}
\usepackage{mathtools}
\usepackage{bbold}
\usepackage{ragged2e}
\usepackage{hyperref}
\usepackage{makecell}

\newlength{\classpageheight}
\newlength{\classpagewidth}
\newcommand*{\citen}[1]{%
  \begingroup
    \romannumeral-`\x 
    \setcitestyle{numbers}%
    \cite{#1}%
  \endgroup
}
\begin{document}
\title{Model Hamiltonian and Time Reversal Breaking Topological Phases
of Anti-ferromagnetic Half-Heusler Materials}
\author{Jiabin Yu}
\affiliation{Department of Physics, the Pennsylvania State University, University Park, PA, 16802}
\author{Binghai Yan}
\affiliation{Department of Condensed Matter Physics, Weizmann Institute of Science, Rehovot, 7610001, Israel}
\author{Chao-Xing Liu}
\email{E-Mail: cxl56@psu.edu}
\affiliation{Department of Physics, the Pennsylvania State University, University Park, PA, 16802}
\begin{abstract}
In this work, we construct a generalized Kane model with a new coupling term between itinerant electron spins
and local magnetic moments of anti-ferromagnetic ordering in order to describe the low energy effective
physics in a large family of anti-ferromagnetic half-Heusler materials.
Topological properties of this generalized Kane model is studied and a large variety of
topological phases, including Dirac semimetal phase, Weyl semimetal phase, nodal line semimetal phase,
type-B triple point semimetal phase,
topological mirror (or glide) insulating phase and anti-ferromagnetic topological insulating phase, are identified
in different parameter regions of our effective models. In particular, we find that
the system is always driven into the anti-ferromagnetic topological insulator phase once a bulk band gap is open,
irrespective of the magnetic moment direction,
thus providing a robust realization of anti-ferromagentic topological insulators.
Furthermore, we discuss the possible realization of these topological phases in realistic anti-ferromagnetic
half-Heusler materials. Our effective model provides a basis for the future study
of physical phenomena in this class of materials.
\end{abstract}
\maketitle
\section{Introduction}
The discovery of time reversal invariant topological insulators
(TIs) \cite{RevModPhys.83.1057,RevModPhys.82.3045} provides us the first example of a novel topological state
that is protected by certain types of symmetry (time reversal symmetry),
and greatly deepens our understanding of the role of symmetry
and topology in electronic band structures of solid materials.
Soon after this discovery, the idea of symmetry protected topological
states is generalized to other systems,
leading to different topological states, including
topological crystalline insulators\cite{PhysRevLett.106.106802,Lin2010,Tanaka2012,xu2012observation,hsieh2012topological} that are protected by
crystalline symmetry, topological superconductors\cite{alicea2012new,schnyder2008classification,lutchyn2010majorana,doi:10.1146/annurev-conmatphys-030212-184337} which
require particle-hole symmetry, and topolgical semimetals (TSMs)\cite{yan2016topological,yang2015weyl,PhysRevB.83.205101,PhysRevX.5.011029,huang2015weyl,xu2015discovery,lv2015experimental,soluyanov2015type,PhysRevB.85.195320,liu2014discovery,PhysRevB.88.125427,PhysRevB.84.235126,bradlyn2016beyond}.
Most current experimental studies of topological states are focused
on non-magnetic materials that preserve time reversal symmetry\cite{yan2012topological}
and ferromagnetic materials (mainly the quantum anomalous Hall effect)\cite{haldane1988model,liu2008quantum,chang2013experimental,yu2010quantized}.
Theoretically, a large variety of topological states can also
exist in materials with other types of magnetic structures, such as
anti-ferromagnetism (AFM) \cite{moore2010,fang2013topological,yoshida2013topological,
wu2015quantum,begue2016identifying,brzezicki2016topological,young2016filling}.
Nevertheless, material proposals of these topological states are still rare.

Half-Heusler compounds are a large group of materials consisting of three metal elements
and have been widely studied for their flexible electronic properties and functionalities \cite{graf2011heusler}.
Around 50 half-Heusler compounds are theoretically predicted to possess inverted band structure
 and can be driven into the topological insulating phase by applying strains
\cite{lin2010half,chadov2010tunable,PhysRevLett.105.096404,PhysRevB.82.125208,yan2014half}.
Unusual surface states were recently observed experimentally in LnPtBi(Ln=Lu, Y) \cite{Liu2016} and LuPtSb \cite{logan2016observation},
and serve as an evidence of non-trivial bulk topology.
Weyl semimetal (WSM) phase was theoretically discussed in several half-Heusler compounds,
including GdPtBi under external magnetic fields\cite{cano2016chiral}
and LaPtBi with in-plane strain\cite{ruan2016symmetry},
and the corresponding evidences were found in recent experiments\cite{Hirschberger2016,shekhar2016observation,suzuki2016large}.
Our interest in this work is focused on possible topological states in
half-Heusler materials with AFM at zero external magnetic field.
AFM has been experimentally observed in RPdBi (R $=$ Er, Ho, Gd, Dy, Tb, Nd) and GdPtBi
\cite{pan2013superconductivity, PhysRevB.84.035208,PhysRevB.90.041109,0953-8984-27-27-275701,nakajima2015topological,pavlosiuk2016antiferromagnetism,pavlosiuk2016magnetic}, with Neel temperature ranging from 1K to 13K.
To describe AFM in half-Heusler compounds, we construct a
generalized six-band Kane model with anti-ferromagnetic coupling terms
based on the symmetry principle and justify this model with the microscopic
tight-binding model. With this model, we predicted a large variety of
time reversal breaking TSM phases, including Dirac semimetal (DSM) phase
(if inversion symmetry breaking is insignificant), WSM phase,
nodal line semimetal (NLSM) phase (if AFM preserves mirror or glide symmetry) and
Type-B triple point semimetal (TPSM) phase\cite{PhysRevX.6.031003} (if AFM preserves $C_{3}$ and glide symmetries),
and topological insulating phases, including
topological mirror (or glide) insulating (TMI) phase
and anti-ferromagnetic topological insulating (AFMTI) phase that is protected by
the combination of time reversal and half translation,
depending on the material parameters.
We also discuss phase diagrams of this model and identify
the candidate half-Heulser compounds to search for these topological phases experimentally.

This paper is organized as follows.
In Sec.\ref{sec:Model Hamiltonian of anti-ferromagnetic half-Heusler materials}, we first present our generalized six-band Kane model
and describe the symmetry aspect of this model.
In Sec.III, we focus on the $4\times 4$ block of this model
which is relatively simple and includes all the four $\Gamma_8$
bands that are close to Fermi energy. A variety of topological phases,
including DSM phase, WSM phase, NLSM phase,
type-B TPSM phase and
TMI phase, were identified for different aligning direction of magnetic moments of AFM. In Sec.IV, we
discuss the limitation of the four-band model and extract the phase diagram of the six-band
Kane model. We demonstrate a robust realization of AFMTI phase
in half-Heusler materials with AFM once anti-ferromagnetic coupling is strong enough. The conclusion is drawn in Sec.V.

\section{Model Hamiltonian of Anti-ferromagnetic Half-Heusler Materials}
\label{sec:Model Hamiltonian of anti-ferromagnetic half-Heusler materials}
\begin{figure}[t]
\includegraphics[width=\columnwidth]{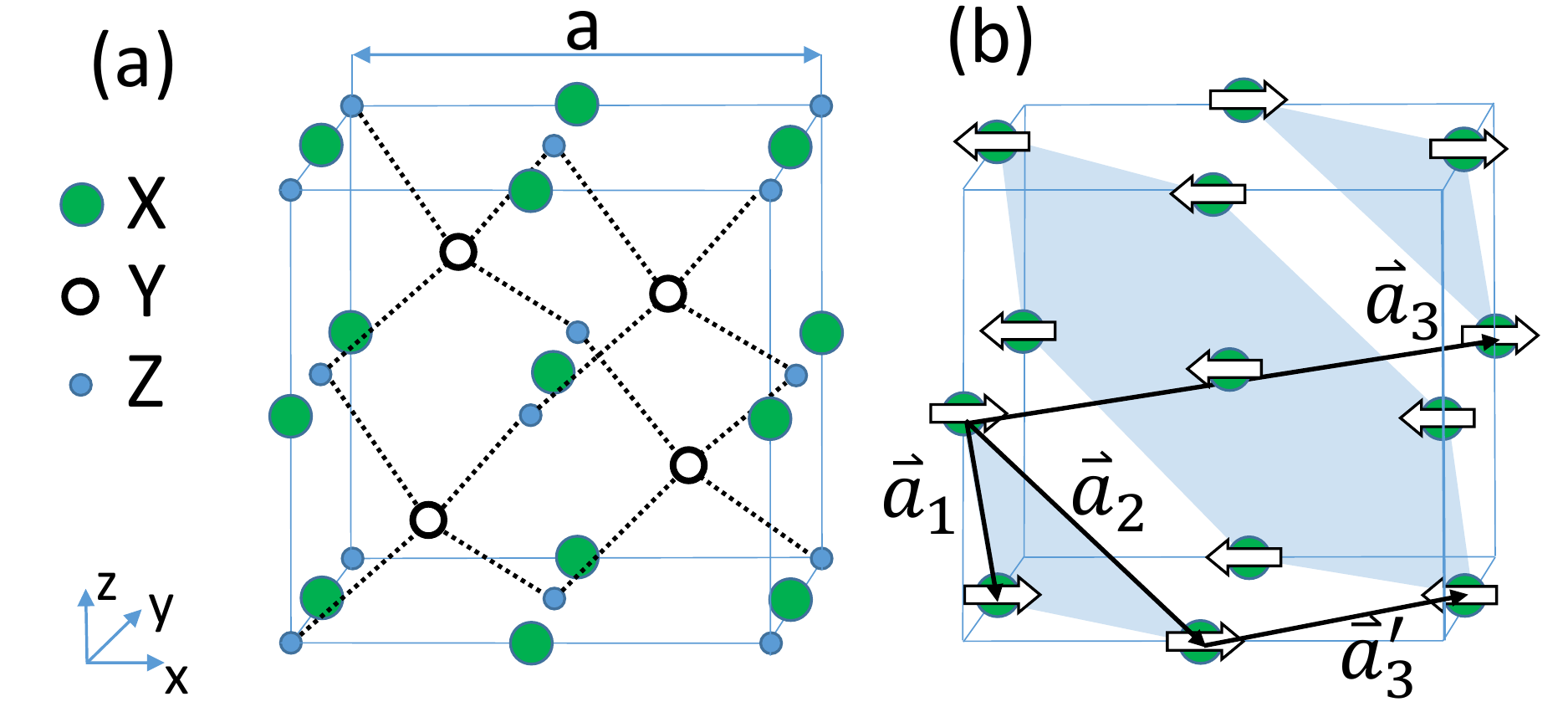}
\caption{
\label{structure}
(a)Crystal structure of half-Heusler alloy XYZ.
(b) The alignment of magnetic moments at X atoms is shown schematically by
arrows.
Magnetic moments align ferromagnetically in one layer perpendicular to (111) direction and anti-ferromagnetically
between two adjacent layers (type-G anti-ferromagnetic ordering).
$\mathbf{a}_1$, $\mathbf{a}_2$ and $\mathbf{a}_3$ label
three lattice basis vectors of the lattice with AFM, whereas $\mathbf{a}_1$, $\mathbf{a}_2$ and $\mathbf{a}'_3$ are
the original lattice basis vectors of FCC lattice without AFM.
}
\end{figure}

We start with a derivation of the model Hamiltonian for
half-Heusler materials with AFM. The half-Heusler compound is normally labelled
by XYZ \cite{canfield1991magnetism},
in which two metal atoms X and Y together play the role of cations
and the metal atom Z is regarded as anions.
The crystal structure of half-Heusler materials is shown in Fig.\ref{structure}a,
in which X and Z atoms form the NaCl-type substructure and
Y and Z atoms form the zinc-blende substructure.
For certain half-Heusler materials with AFM, including GdPtBi, DyPdBi, HoPdBi and TbPdBi
\cite{pavlosiuk2016antiferromagnetism,nakajima2015topological,PhysRevB.90.041109},
magnetic moments come from the X atom and
align ferromagnetically in one layer perpendicular to (111) direction and anti-ferromagnetically
between two adjacent layers (type-G anti-ferromagnetic ordering), as shown in Fig.\ref{structure}b,
while the exact direction of magnetic moments
is still not clear and may depend on detailed compound composition.

To construct the effective Hamiltonian,
we need to understand the symmetry aspect of this crystal. Without magnetic moments,
the space group of half-Heusler compounds is F$\bar{4}3$m \cite{canfield1991magnetism}
and the corresponding point group is $T_d$, similar to zinc-blende structure.
As a result, the lattice vectors should be chosen to be those of a face-centered cubic
lattice, labelled as $\mathbf{a}_1$, $\mathbf{a}_2$ and $\mathbf{a'}_3$ in Fig.\ref{structure}b.
The existence of type-G AFM has two main effects to the symmetry of the system:
(1) it doubles the lattice vector and
the unit-cell along the $\mathbf{a'}_3$ direction,
and the new lattice vector is labeled as $\mathbf{a}_3=2\mathbf{a'}_3$;
(2) in the mean field level, the magnetic moments of AFM give rise to exchange coupling to electron spins.
Due to the doubling of unit-cell, the space group and point group of the crystalline structure are reduced to
$R3m$\cite{li2015electronic} and $C_{3v}$, where the latter has two generators: three-fold rotation
$C_3{(111)}$ along (111) direction
and mirror symmetry $\pi_{(1\bar{1}0)}$ with respect to $(1\bar{1}0)$ plane.
As a pseudo-vector, magnetic moments of AFM can further lower the symmetry.
Besides the above crystal symmetries,
another essential symmetry due to AFM is the combination
of time reversal $\Theta$ and half tranlation $T_{\mathbf{a}'_3}$,
denoted as $S\equiv T_{\mathbf{a}'_3}\Theta$.

Similar to the semiconductors with the zinc-blende structure,
the low energy physics of half-Heusler materials can be described by
two $\Gamma_6$ bands with s-orbital nature, labelled by
$|\Gamma_6,S_z\rangle$ with z-directional spin component $S_z=\pm\frac{1}{2}$,
and four $\Gamma_8$ bands with p-orbital nature,
labelled by $|\Gamma_8,J_z\rangle$ with total angular momentum $J_z=\pm\frac{1}{2},
\pm\frac{3}{2}$ which is a combination of spin and orbital angular momentum.
Here the notations $\Gamma_6$ and $\Gamma_8$ refer to the irreducible representations
of the corresponding bands under the $T_d$ group \cite{winkler2003spin} and we still keep
this notation even though the symmetry is lowered.
The form and symmetry of the basis wave functions are also justified
with microscopic local atomic orbitals, as discussed in details
in Appendix \ref{app:Tight-binding}.
It should be emphasized that due to the doubling of the unit-cell, the
basis wave functions in the new unit-cell is a bonding or anti-bonding
state of the basis functions in the original unit-cell. For the low energy physics,
we find both $|\Gamma_6,S_z\rangle$ and $|\Gamma_8,J_z\rangle$ are bonding states
in the doubled unit-cell and are still suitable to be bases.
To obtain a matrix expansion of the effective Hamiltonian, we need to classify
all the matrices according to the irreducible representations of the symmetry
group for the above basis.
The Hamiltonian should be invariant
under symmetry operations of the symmetry group, which means it belongs to the identity
representation, and can be constructed by combining
the symmetrized matrices and symmetrized tensor components of physical quantities
(such as the momentum ${\bf k}$ and the order parameters ${\bf M}$ for AFM)
which belong to the same irreducible representation.\cite{winkler2003spin}
This symmetry
approach allows us to systematically construct the full Hamiltonian, including
the standard six-band Kane model and an additional AFM term.
The standard six-band Kane model is \cite{winkler2003spin}
\begin{equation}
H_{Kane}=
\left(
\begin{array}{c|c}
H_{\Gamma_6}(\mathbf{k})&V\\
\hline
V^{\dagger}&H_{\Gamma_8}(\mathbf{k})\\
\end{array}
\right),
\end{equation}
where
\begin{equation}
\begin{array}{l}
V=\frac{P}{\sqrt{6}}\left(\begin{matrix}
-\sqrt{3}k_+ & 2k_z & k_- & 0 \\
0 & -k_+ & 2k_z & \sqrt{3}k_-\end{matrix}
\right)\\
+\frac{B^+_{8v}}{\sqrt{6}}\left(\begin{matrix}
\sqrt{3}k_- k_z & 2 i k_x k_y & k_+ k_z & 0 \\
0 & k_- k_z & 2 i k_x k_y & \sqrt{3}k_+ k_z\end{matrix}
\right)\\
+\frac{B^-_{8v}}{3\sqrt{2}}\left(\begin{matrix}
0 & \sqrt{3}K^2 &0 & k_{\parallel}^2-2k_z^2 \\
-k_{\parallel}^2+2k_z^2 & 0 & -\sqrt{3}K^2 & 0\end{matrix}
\right)
\end{array},
\end{equation}
\begin{equation}
H_{\Gamma_6}(\mathbf{k})=(E_c+\beta_c k^2)\mathbb{1}_{2}
\end{equation}
with $\mathbb{1}_{2}$ to be the $2\times 2$ identity matrix,
\begin{equation}
H_{\Gamma_8}(\mathbf{k})=H_0(\mathbf{k})+H_C(\mathbf{k})
\end{equation}
with
\begin{equation}
\begin{array}{c}
H_{0}(\mathbf{k})=
\frac{4}{15}(J_x^2+J_y^2+J_z^2) h_{0}+\frac{1}{3}(2J_z^2-J_x^2-J_y^2) h_{1}\\
+\frac{1}{\sqrt{3}}(J_x^2-J_y^2) h_{2}+\frac{2}{\sqrt{3}}J_{xy} h_{3}+\frac{2}{\sqrt{3}}J_{zx} h_{4}+\frac{2}{\sqrt{3}}J_{yz} h_{5}
\end{array}
\end{equation}
 and
\begin{equation}
H_C(\mathbf{k})=\frac{2}{\sqrt{3}}C (k_x V_x+ k_y V_y+ k_z V_z).
\end{equation}
Here we have
$h_0=E_v-\beta_c\gamma_1 k^2$,
$h_{1}=\beta_c\gamma_2 (2 k_z^2-k_{\parallel}^2)$,
$h_{2}=\sqrt{3}\beta_c\gamma_2 K^2$,
$h_{3}=2\sqrt{3}\beta_c \gamma_3 k_x k_y$,
$h_{4}=2\sqrt{3}\beta_c \gamma_3 k_x k_z$,
and
$h_{5}=2\sqrt{3}\beta_c \gamma_3 k_y k_z$.
$J_i$'s are angular momentum matrices for spin $3/2$ (see Appendix \ref{app:irrep_C3v} for details),
$J_{ij}=\frac{1}{2}\{J_i,J_j\}$,
$V_x=\frac{1}{2}\{J_x,J_y^2-J_z^2\}$,
$V_y=\frac{1}{2}\{J_y,J_z^2-J_x^2\}$,
$V_z=\frac{1}{2}\{J_z,J_x^2-J_y^2\}$,
$\beta_c=\hbar^2/(2 m')$,
$m'$ is the effective mass of $\Gamma_6$ bands near $\Gamma$ point,
$k^2=k_x^2+k_y^2+k_z^2$,
$k_{\parallel}^2=k_x^2+k_y^2$,
$K^2=k_x^2-k_y^2$
and $k_{\pm}=k_x\pm i k_y$.
In this work, $\beta_c > 0$, $\gamma_2\neq 0 $ and $\gamma_3 \neq 0$
are always assumed, unless being specified otherwise.
Besides the above terms in the standard Kane model, AFM can lead to new terms,
which can be constructed in a similar manner with the symmetry group combining point group
$C_{3v}$, $T_{\mathbf{a}'_3}$ and $\Theta$ (see Appendix \ref{app:irrep_C3v} for more details).
For the materials that we are interested in,
the main influence of AFM only occurs for the four $\Gamma_8$ bands.
Thus, we only consider the anti-ferromagnetic terms in the basis of the four $\Gamma_8$ bands
and the corresponding Hamiltonian is given by
\begin{equation}
\arraycolsep=1pt\def\arraystretch{1.2}
\begin{array}{l}
	H_{AFM}=
\frac{4}{15}(J_x^2+J_y^2+J_z^2) \xi_{0}+\frac{1}{3}(2J_z^2-J_x^2-J_y^2) \xi_{1}\\
+\frac{1}{\sqrt{3}}(J_x^2-J_y^2) \xi_{2}+\frac{2}{\sqrt{3}}J_{xy} \xi_{3}+\frac{2}{\sqrt{3}}J_{zx} \xi_{4}+\frac{2}{\sqrt{3}}J_{yz} \xi_{5},\\
\end{array}\label{eq:Ham_AFM}
\end{equation}
where the detailed expressions for $\xi_i$ ($i=0,\dots,5$) are listed in Tab.\ref{tab:xi_M}.
\begin{table}
\begingroup\makeatletter\def\f@size{8}\check@mathfonts
$$
\begin{array}{|c|c|}
\hline
\xi_0 &\alpha_1(M_x^2+M_y^2+M_z^2)+\beta_1(M_y M_z+M_zM_x+M_xM_y)\\
\hline
\xi_1 &\alpha_3(2 M_z^2-M_x^2-M_y^2)+\beta_3(2M_xM_y-M_yM_z-M_zM_x)\\
\hline
\xi_2 &\alpha_3\sqrt{3}(M_x^2-M_y^2)+\beta_3\sqrt{3}(M_yM_z-M_xM_z)\\
\hline
\xi_3 &
\begin{matrix}
\alpha_2(M_x^2+M_y^2+M_z^2)+\beta_2(M_yM_z+M_zM_x+M_xM_y)\\
+2[\alpha_4(2 M_z^2-M_x^2-M_y^2)+\beta_4(2M_xM_y-M_yM_z-M_zM_x)]\\
\end{matrix}\\
\hline
\xi_4 &
\begin{matrix}
\alpha_2(M_x^2+M_y^2+M_z^2)+\beta_2(M_yM_z+M_zM_x+M_xM_y)\\
-[\alpha_4(2 M_z^2-M_x^2-M_y^2)+\beta_4(2M_xM_y-M_yM_z-M_zM_x)]\\
-\sqrt{3}[\alpha_4\sqrt{3}(M_x^2-M_y^2)+\beta_4\sqrt{3}(M_yM_z-M_xM_z)]\\
\end{matrix}\\
\hline
\xi_5 &\begin{matrix}
\alpha_2(M_x^2+M_y^2+M_z^2)+\beta_2(M_yM_z+M_zM_x+M_xM_y)\\
-[\alpha_4(2 M_z^2-M_x^2-M_y^2)+\beta_4(2M_xM_y-M_yM_z-M_zM_x)]\\
+\sqrt{3}[\alpha_4\sqrt{3}(M_x^2-M_y^2)+\beta_4\sqrt{3}(M_yM_z-M_xM_z)]\\
\end{matrix}\\
\hline
\end{array}
$$
\endgroup
\caption{
\label{tab:xi_M}
Expressions for the $\xi_i$ parameters ($i=0,1,2,3,4,5$) in the AFM Hamiltonian
(Eq.\ref{eq:Ham_AFM}).
Here $M_i$'s are three components of the anti-ferromagnetic magnetic moment, and
$\alpha_i$'s, $\beta_j$'s are parameters independent of each other and momentum $\mathbf{k}$.}
\end{table}

Next we discuss the symmetry properties of this Hamiltonian.
For the standard Kane model $H_{Kane}$, if the parameter $C$ in $H_C$ is zero,
the Hamiltonian possess $O_h$ point group.
Non-zero $H_C$ term breaks inversion symmetry and lowers the point group
from $O_h$ to $T_d$.
The existence of AFM results in the doubling of the unit cell along $\mathbf{a}'_3$
and reduce the point group of the lattice to $C_{3v}$.
For a fixed non-zero anti-ferromagnetic order ${\bf M}$, the AFM term $H_{AFM}$ will further reduce the symmetry.
If $\mathbf{M}$ is along the (111) direction,
$C_3(111)$ symmetry is maintained but all mirror symmetries in $C_{3v}$ are broken,
whereas the glide symmetries $T_{\mathbf{a}'_3}\pi_{(1\bar{1}0)}$, $T_{\mathbf{a}'_3}\pi_{(\bar{1}01)}$ and $T_{\mathbf{a}'_3}\pi_{(01\bar{1})}$ are preserved.
If $\mathbf{M}$ lies in a mirror plane $\alpha$ ($\alpha=(1\bar{1}0)$ or $(\bar{1}01)$ or $(01\bar{1})$)
but away from the (111) direction,
all symmetries in $C_{3v}$ are broken,
whereas the glide symmetry $T_{\mathbf{a}'_3}\pi_{\alpha}$ is preserved.
If $\mathbf{M}$ is perpendicular to a mirror plane of the lattice,
the only remaining symmetry is that mirror symmetry.
For a generic AFM Hamiltonian, $\mathbf{M}$ will break all the symmetries in $C_{3v}$ group, as well
as the combination with $T_{\mathbf{a}'_3}$.
Furthermore, we notice that only quadratic terms of $\mathbf{M}$ appear
in our Hamiltonian while any linear $\mathbf{M}$ terms vanish.
This is because $\mathbf{M}$ reverses its sign under translation $T_{\mathbf{a}'_3}$,
while translation $T_{\mathbf{a}'_3}$ is just identity matrix for the basis of four $\Gamma_8$ bands
and thus commutes with any representation matrix (see Appendix \ref{app:irrep_C3v} for more details).
This suggests that any term with the odd orders of the anti-ferromagnetic order parameter $\mathbf{M}$ cannot exist.

\section{Topological Phases in the Four-band Model}
Since we are interested in the half-Heusler materials with inverted
band structures, only four $\Gamma_8$
bands appear near the Fermi energy while the $\Gamma_6$ bands are far below
the Fermi energy. Thus, we first focus on the four $\Gamma_8$ bands
with the Hamitonian $H_{\Gamma_8}(\mathbf{k})+H_{AFM}$.
For the inverted band structure, the Fermi energy is between the $\Gamma_8$ bands,
and the two $\Gamma_8$ bands with lower energies are valence bands while
the other two $\Gamma_8$ bands with higher energies are conduction bands.
We emphasize that the $\Gamma_6$ bands are
important for certain types of topological states even though they are
away from the Fermi energy, as discussed in details in the next section.
However, for the TSM phases discussed in this section,
only the $\Gamma_8$ bands are essential. Another advantage of the
4-band Hamiltonian is that it can be solved analytically in certain limit,
thus providing us valuable insight into the underlying physics.
In this section, we first focus on the case without inversion symmetry
breaking term (i.e. choosing $H_C(\mathbf{k})=0$) and reveal the occurence of DSM
phase due to the coexistence of inversion symmetry and anti-unitary
$S$ symmetry. In realistic system, inversion symmetry
is broken and DSM phase becomes unstable. Nevertheless,
DSM phase can be viewed as the ``parent'' phase
to generate other TSM phases after
including $H_C(\mathbf{k})$.
We further study the situation with non-zero inversion symmetry breaking term $H_C(\mathbf{k})$,
focusing on the situations with
(1) magnetic moments polarized within or perpendicular to the $(1\bar{1}0)$ plane
and (2) $\mathbf{M}$ along the (111) direction.

\subsection{Dirac Semimetal Phase and Topological Mirror Insulating Phase}
\label{sec:Dirac Semimetal Phase and Topological Mirror Insulating Phase}

\begin{figure}[t]
\includegraphics[width=\columnwidth]{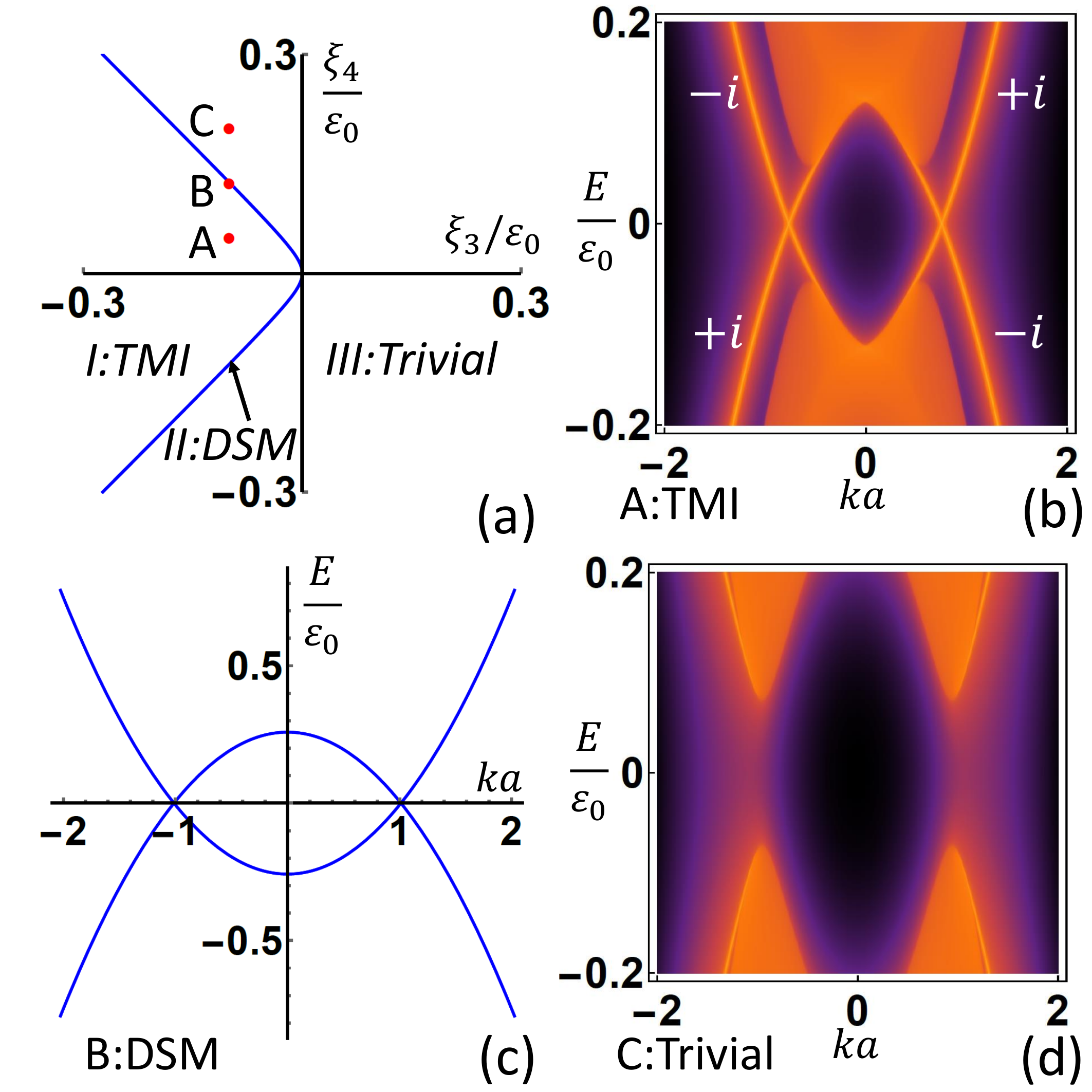}
\caption{
\label{Dirac}
(a) Phase Diagram as a function of $\xi_3$ and $\xi_4$ for the Hamiltonian
$H_0(\mathbf{k})+H_{AFM}$.
Here phase I is TMI phase,
phase III is trivial insulating phase, and phase II is DSM phase
that serves as a critical line separating phase I and III.
(b) reveals the local density of states on (001) surface for point A in (a) with $(\xi_3,\xi_4)=(-0.1,0.05)\varepsilon_0$, which is in TMI phase. $k$ is along (110) direction. $\pm i$ are mirror eigenvalues of corresponding bands.
(c) Energy spectrum for the point B in (a), which is in DSM phase.
$k$ is along the line connecting two Dirac points and each band is doubly degenerated.
(d) reveals the local density of states on (001) surface for point C in (a) with $(\xi_3,\xi_4)=(-0.1,0.2)\varepsilon_0$, which is in the trivial insulating phase. $k$ is along (110) direction.}
\end{figure}

In this part, $H_C(\mathbf{k})=0$ is assumed and the total Hamiltonian takes the form
$H_0(\mathbf{k})+H_{AFM}$ with inversion symmetry.
In this case, the eigenenergy of Hamiltonian can be solved
analytically as
\begin{equation}
\label{eq:0+AFM}
E_{0+AFM,\pm}(\mathbf{k})=(h_0+\xi_0)\pm\sqrt{\sum_{i=1}^5(h_i+\xi_i)^2}.
\end{equation}
Without AFM term, $\xi_i=0$ for $i=0,\dots,5$ in Eq.\ref{eq:Ham_AFM},
and there are four-fold degeneracy of the $\Gamma_8$ bands at the $\Gamma$ point ($\mathbf{k}=0$) due to
the $T_d$ group symmetry. The conduction and valence bands touch each other quadratically at $\mathbf{k}=0$,
leading to a critical semimetal phase for the four-band Luttinger model $H_0(\mathbf{k})$.
Early studies have demonstrated various TSM phases induced by strain or external magnetic fields
in this system \cite{shekhar2016observation,ruan2016symmetry}.
The AFM term $H_{AFM}$ can lower the symmetry of the system and remove
the four-fold degeneracy at $\Gamma$ point.
However, since $H_{AFM}$ preserves the inversion symmetry and $S$ symmetry, all the bands
are still doubly degenerate, similar to the Kramer's degeneracy due to inversion and time reversal symmetries\cite{1367-2630-9-9-356}.

Next we will study the influence of AFM term on the energy dispersion.
The AFM term can lead to a non-zero gap at $\Gamma$ point, given by $2\sqrt{\xi_1^2+\xi_2^2+\xi_3^2+\xi_4^2+\xi_5^2}$.
Thus, the two doubly degenerate bands are split at the $\Gamma$ point, but they
may cross each other at some finite momenta ${\bf k}$, giving rise to a semimetal phase.
The realization of such semimetal phase requires $h_i+\xi_i=0$ for all $i=1,\dots,5$
and the details are discussed in Appendix \ref{app:gapless_conditions_4band0}.
Here we focus on the cases where magnetic moments of AFM
are perpendicular to $(1\bar{1}0)$ plane
($M_x=-M_y$ and $M_z=0$, where $M_{x,y,z}$ are x,y,z components of AFM magnetic moment $\mathbf{M}$) with mirror symmetry $\pi_{(1\bar{1}0)}$, or lie in the $(1\bar{1}0)$ plane ($M_x=M_y$)
with the glide symmetry $T_{\mathbf{a}'_3}\pi_{(1\bar{1}0)}$.
According to the expressions of $\xi_i$'s,
both conditions imply the same requirement for parameters: $\xi_2=0$ and $\xi_4=\xi_5$,
which is reasonable since $\pi_{(1\bar{1}0)}$ and
$T_{\mathbf{a}'_3}\pi_{(1\bar{1}0)}$
have the same matrix representation for four basis wave functions of $\Gamma_8$ bands.
For the existence of gapless points,
one of the following additional conditions is required for the values of $\xi_1$, $\xi_3$ and $\xi_4$
(see Appendix C for more details):
(i) $\xi_3\beta_c\gamma_3<0$, $\sqrt{3}
\gamma_3 \xi_3\xi_1+\gamma_2 \left(\xi_3^2 -\xi_4^2\right)=0$;
(ii) $\xi_3=\xi_4=0$, $\xi_1 \beta_c \gamma_2<0$;
(iii) $\xi_4=0$, $\xi_3\beta_c\gamma_3>0$, $\sqrt{3}\gamma_3\xi_1-\gamma_2\xi_3=0$.
Due to $\xi_2=0$ and $\xi_4=\xi_5$, we require $k_x^2=k_y^2$ and $k_x k_z= k_y k_z$ from $h_i=-\xi_i$ with $i=2,4,5$
, indicating two possibilities for
the locations of gapless point, either on the plane $(1\bar{1}0)$ with the form $(k_1,k_1,k_3)$
for the conditions (i) and (ii),
or on the $(1,-1,0)$ axis with the form $(k_1,-k_1,0)$ for the condition (iii).
Due to the $S$ symmetry, if a gapless point occurs at a finite momentum ${\bf k}_0$,
there must be another one at $-{\bf k}_0$, leading to even number of
gapless points. The number of gapless points are confirmed to be 2 by solving for positions of gapless points in each case (see Appendix \ref{app:gapless_conditions_4band0} for more details).

Based on the above conditions for gapless points, we can further extract the phase diagram of this model
as a function of $(\xi_3,\xi_4)$.
An example of a phase diagram is shown in Fig.\ref{Dirac}a for the  choices of parameters listed in
Tab.\ref{tab:Diracfig} in Appendix \ref{app:parameters}.
The blue line in the phase diagram, labelled by II, represents DSM phase.
For our choices of the parameters, the condition (i) or (ii) can be satisfied
and Dirac points are on the $(1\bar{1}0)$ plane.
Fig.\ref{Dirac}c reveals a typical energy dispersion of the semimetal phase at the point B in Fig.\ref{Dirac}a
with $\xi_3/\varepsilon_0=-\frac{1}{10}$ and $\xi_4/\varepsilon_0=\frac{1}{50} \sqrt{8 \sqrt{3}+25}$ to satisfy the condition (i), where the energy unit $\varepsilon_0$ is defined as $\varepsilon_0\equiv\beta_c/a^2$ and $a$ is a real positive parameter with unit of length.
The energy dispersion around the gapless point behaves linearly, thus forming two Dirac cones
at ${\bf k}a=\pm \frac{1}{\sqrt[4]{12}}(1,1,-\frac{1}{5} \sqrt{8 \sqrt{3}+25})$, given the double degeneracy for each band.
Further theoretical analysis of the effective low-energy Hamiltonian
expanded around these two gapless points confirms this DSM phase, as
shown in details in Appendix \ref{app:verification_DP}.

The DSM phase separates
two insulating phases, labeled by I and III in Fig.\ref{Dirac}a. To identify the nature of these two insulating phases,
we perform a numerical calculation of energy dispersion on the $(001)$ surface of an approximately semi-infinite sample.\cite{sancho1985highly}
The local density of states at the top surface along $k_x=k_y$ axis is shown in Fig.\ref{Dirac}b for the point A with $(\xi_3,\xi_4)=(-0.1,0.05)\varepsilon_0$ in the
phase I and Fig.\ref{Dirac}d for the point C with $(\xi_3,\xi_4)=(-0.1,0.2)\varepsilon_0$ in the phase III, respectively.
One can see two sets of  gapless modes appearing for the phase I
while a full gap existing for the phase III.
Thus, we expect the phase I is topologically non-trivial while the phase III
is trivial.
We also perform a calculation of mirror Chern number (MCN) \cite{PhysRevB.78.045426} on the mirror or glide plane
($1\bar{1}0$) for this system and find MCN to be 2 for the phase I and 0 for the phase III.
This confirms that two sets of gapless modes in Fig.\ref{Dirac}b are protected by mirror or glide symmetry and makes phase I to be TMI phase.
Thus, DSM phase can be viewed as the topological phase
transition point between a TMI phase and a trivial insulating phase.

We emphasize that, in realistic half-Heusler materials, inversion symmetry is absent and
the Kramer's degeneracy at a generic $\mathbf{k}$ is split for both the conduction and valence bands,
which means Dirac points are also split.
However, DSM phase will evolve into other TSM phases,
as discussed in the next section. Therefore, DSM phase can be viewed as
the ``parent'' phase to search for and understand other topological phases.

\subsection{Weyl Semimetal Phase and Topological Mirror Insulating Phase}
\begin{figure}[t]
\includegraphics[width=\columnwidth]{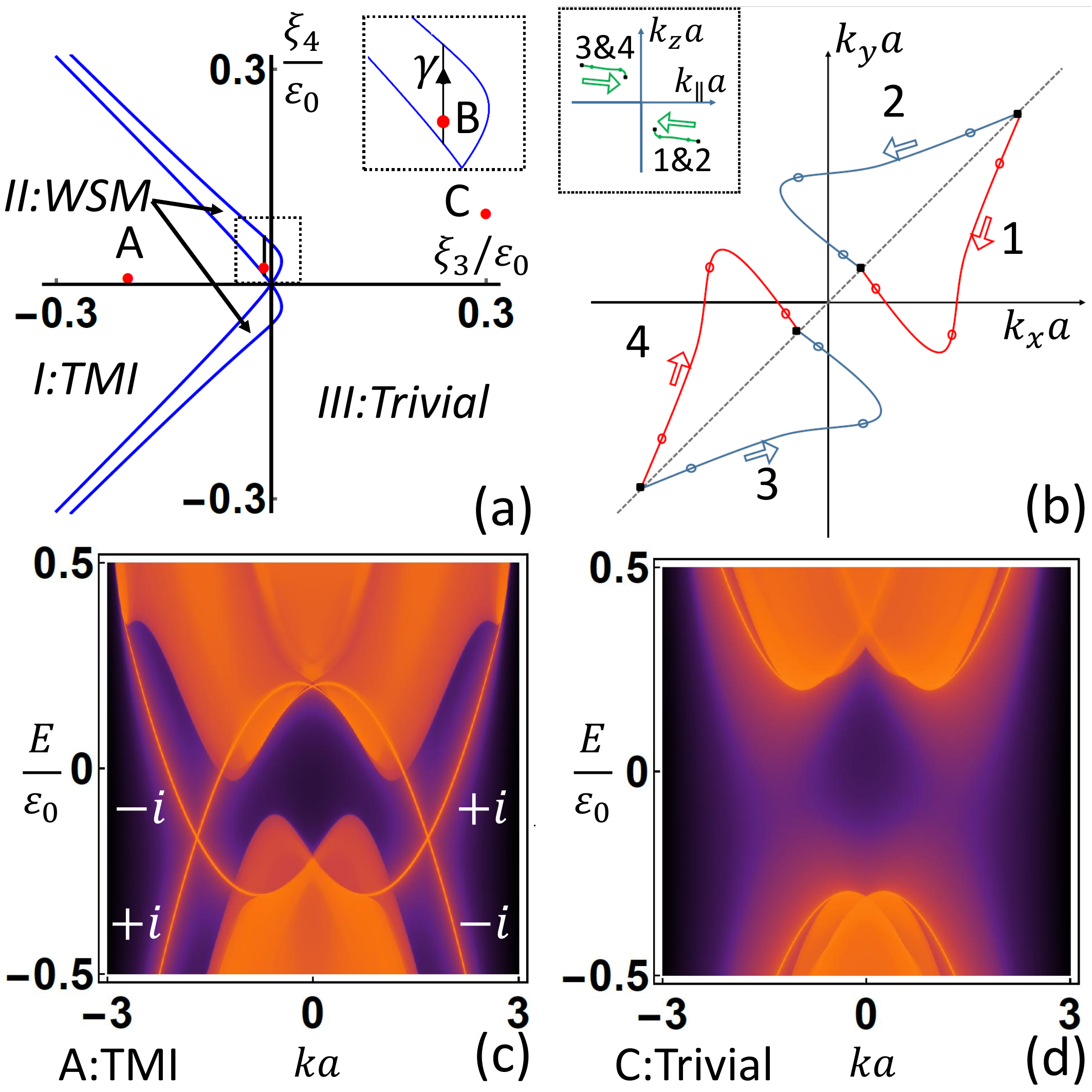}
\caption{
\label{WeylPD}
(a)Phase Diagram as a function of $\xi_3$ and $\xi_4$, in which the phase I is a topological mirror insulator,
the phase III is a trivial insulator, and the phase II is a Weyl semi-metal phase.
The determination of boundary of phase II is described in Appendix \ref{app:WSM_critical_line}.
The inset is a zoom-in version for the boxed part of phase II.
(b) is a schematic for the motion of Weyl points along the path $\gamma$ in the inset of (a).
The main graph shows the projection of the motion on the $k_x-k_y$ plane, and the inset depicts the
projection of the motion on $(1\bar{1}0)$ plane which is the mirror or glide plane.
(c) and (d) reveal the local density of states on (001) surface for point A ($\xi_3/\varepsilon_0=-0.2$ and $\xi_4/\varepsilon_0=0.01$) and C ($\xi_3/\varepsilon_0=0.3$ and $\xi_4/\varepsilon_0=0.1$) in (a) respectively. $k$ is chosen to be along (110) direction. $\pm i$ in (c) are mirror eigenvalues of corresponding bands. A is in TMI phase while C is in trivial insulating phase.}
\end{figure}

In this part, we include the inversion symmetry breaking term $H_C({\bf k})$ into the four-band Hamiltonian and the total Hamiltonian becomes $H_{\Gamma_8}(\mathbf{k})+H_{AFM}$.
As a consequence, the Kramer's degeneracy of each band at a generic momentum $\mathbf{k}$ is split.
We still consider the magnetic moments of anti-ferromagnetic ordering aligning within or perpendicular to
the $(1\bar{1}0)$ plane to preserve either the glide symmetry $T_{\mathbf{a}'_3}\pi_{(1\bar{1}0)}$ or the mirror symmetry $\pi_{(1\bar{1}0)}$, which gives $\xi_2=0$ and $\xi_4=\xi_5$.
Due to the existence of $H_C$ term, the full Hamiltonian cannot be diagonalized analytically
and thus numerical methods are adopted to extract phase diagram.
All the parameters are the same as previous choices,
except the $C$ parameter which is chosen as $C=0.2(\beta_c/a)$, as listed in Tab.\ref{tab:Weylfig}
in Appendix \ref{app:parameters}.

The phase diagram as a function of $\xi_3$ and $\xi_4$ is shown in Fig.\ref{WeylPD}a.
The phases I and III in Fig.\ref{WeylPD}a remain robust due to existence of mirror or glide symmetry.
A direct calculation
of surface local density of states, as well as MCN, shows four surface modes
with MCN being 2 (Fig.\ref{WeylPD}c) for point A in Fig.\ref{WeylPD}a
($\xi_3/\varepsilon_0=-0.2$ and $\xi_4/\varepsilon_0=0.01$) and a full gap with zero MCN (Fig.\ref{WeylPD}d) for point C in Fig.\ref{WeylPD}a ($\xi_3/\varepsilon_0=0.3$ and $\xi_4/\varepsilon_0=0.1$).

\begin{figure}[t]
\includegraphics[width=\columnwidth]{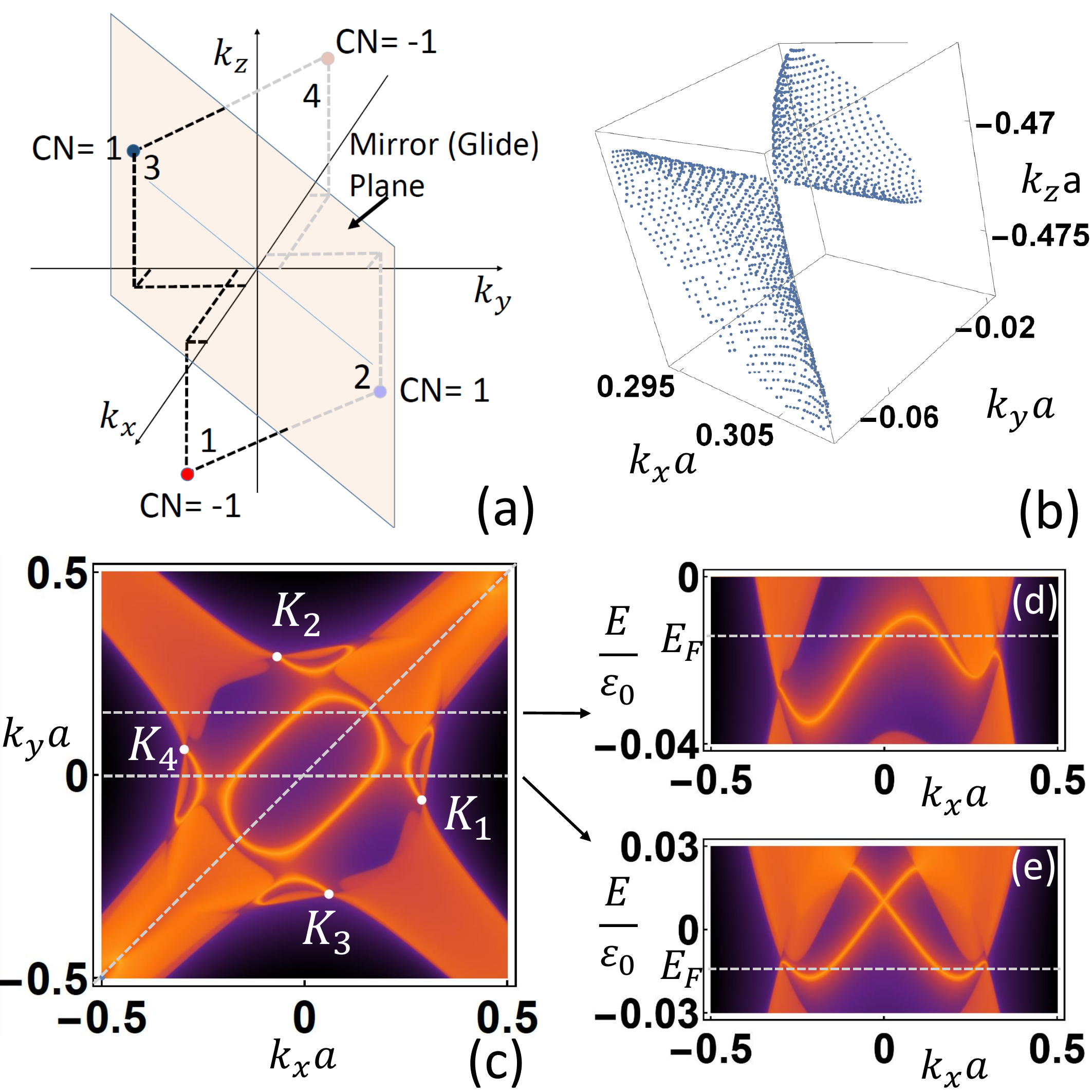}
\caption{
\label{WeylSE}
This figure is for point B in Fig.\ref{WeylPD}a with $\xi_3=-0.01\varepsilon_0$ and $\xi_4=0.025\varepsilon_0$.
(a) shows schematically the positions of four Weyl points with the corresponding CNs.
(b)The Fermi surface around Weyl point $\mathbf{K}_1$ shows that it is a type-II Weyl point.
(c)Local density of states on (001) surface at the energy of Weyl points (Fermi energy).
Weyl points are represented by white dots and connected to Fermi arcs.
Two horizontal grey dashed lines correspond to the momentum line $k_y a=0$ and $k_y a=0.15$.
The grey dashed line along the diagonal direction corresponds to the momentum line $k_x=k_y$ with
mirror or glide symmetry.
(d) and (e) depict local density of states on (001) surface along the momentum line $k_y a =0.15$ and $k_y a=0$ respectively, and the horizontal lines inside them mark the chosen Fermi energy (energy of Weyl points).
}
\end{figure}

We notice that the phase II is expanded from a line of Dirac semimetal phase in Fig.\ref{Dirac}a
to a region of WSM phase in Fig.\ref{WeylPD}a.
The reason is that $H_C$ breaks the inversion symmetry and splits each Dirac cone in Dirac semimetal phase into two Weyl points.
Since there are two Dirac points in the DSM phase,
the phase II in Fig.\ref{WeylPD}a typically has four Weyl points in the whole momentum space,
denoted as ${\bf K}_i$ ($i=1,2,3,4$), as shown in Fig.\ref{WeylPD}b and Fig.\ref{WeylSE}a.
Different Weyl points can be related to each other by symmetries:
$\mathbf{K}_1$ and $\mathbf{K}_2$ (or $\mathbf{K}_3$ and $\mathbf{K}_4$) are related by
$\Pi_{(1\bar{1}0)}\equiv\pi_{(1\bar{1}0)}$ or $T_{\mathbf{a}'_3}\pi_{(1\bar{1}0)}$ and therefore dubbed a `$\Pi$ pair',
$\mathbf{K}_1$ and $\mathbf{K}_3$ (or $\mathbf{K}_2$ and $\mathbf{K}_4$) are related by
$\Pi_{(1\bar{1}0)}S$ and called a `$\Pi S$ pair', and finally
$\mathbf{K}_1$ and $\mathbf{K}_4$ (or $\mathbf{K}_2$ and $\mathbf{K}_3$) are related by
$S$ and named a `$S$ pair'. It is known \cite{volovik2003universe,volovik2007quantum,volovik2013topology} that a Weyl fermion can carry topological charge
or chirality, which can be extracted from Chern number (CN) on a small spherical surface surrounding the Weyl point.
Two Weyl points related by mirror symmetry have opposite Chern numbers while time reversal operation leaves
Chern number of a Weyl point unchanged. As a result, a $\Pi$ pair carries opposite CNs
and so does a $\Pi S$ pair, while a $S$ pair has the same CN.
Due to the existence of topological charge, a single Weyl point cannot be gaped
and it can only move in the momentum space when tuning the parameters until it merges with another Weyl point with opposite CN.
We track the motion of Weyl points through the path $\gamma$ in the inset of Fig.\ref{WeylPD}a,
and find $\mathbf{K}_1$ and $\mathbf{K}_2$ (or $\mathbf{K}_3$ and $\mathbf{K}_4$)
emerge from a point in the mirror (or glide) plane as a $\Pi$ pair and move in the momentum space
and finally annihilate at another point on the mirror (or glide) plane, as shown in Fig.\ref{WeylPD}b with its inset.

Now we focus on the energy dispersion of the WSM phase by choosing
the B point in Fig.\ref{WeylPD}a with $\xi_3=-0.01\varepsilon_0$ and $\xi_4=0.025\varepsilon_0$ as an example.
Four Weyl points are approximately located at
$$
\begin{array}{l}
\mathbf{K}_1 a=\{0.30,-0.047,-0.47\},\\
\mathbf{K}_2 a=\{-0.047,0.30,-0.47\},\\
\mathbf{K}_3 a=\{0.047,-0.30,0.47\},\\
\mathbf{K}_4 a=\{-0.30,0.047,0.47\}
\end{array}
$$
in the momentum space, respectively.
By integrating the Berry curvature on the sphere around each point,
we found CNs of four Weyl points are $CN_1=-1$,$CN_2=1$,$CN_3=1$ and $CN_4=-1$ (see Fig.\ref{WeylSE}a),
which is consistent with the symmetry analysis above.
The Fermi surface around $\mathbf{K}_1$ at the energy of the Weyl point, as shown in Fig.\ref{WeylSE}b,
demonstrates the existence of a type II Weyl point in our system.
Type-II Weyl fermions are topologically non-trivial and can lead to Fermi arc \cite{soluyanov2015type}
on the surface.
Thus, we perform a calculation of local density of states at the energy of Weyl points
on the $(001)$ surface of an approximately semi-infinite sample\cite{sancho1985highly}, as shown in Fig.\ref{WeylSE}c,
in which a complex surface Fermi arcs overlaps with bulk bands.
Weyl points are depicted by white points in the plot and there is one Fermi arc
starting from each Weyl point and merging into bulk bands.
Additional surface states with same energy exist and they form a circle surrounding the $\Gamma$ point ${\bf k}=0$.
The surface energy dispersions along the momentum lines $k_y a=0$ and $k_y a=0.15$
are shown in Fig.\ref{WeylSE}d and  Fig.\ref{WeylSE}e.
We notice one chiral edge mode existing along the momentum line $k_y a=0.15$
while a helical edge mode along the line $k_y a=0$.
We may treat $k_y$ as a parameter and consider
two-dimensional (2D) planes formed by $k_z$ and $k_x$ for different $k_y$.
According to the bulk-edge correspondence,
the existence of chiral edge mode for $k_y a=0.15$ at the surface suggests $CN=-1$ for the
corresponding 2D plane. Similarly, CN for the 2D plane at $k_y a=0$ should be zero.
This is consistent with the fact that two 2D planes at $k_y a=0.15$
and $k_ya=0$ enclose one Weyl point $\mathbf{K}_4$ whose CN is $-1$.
However, the helical edge mode along the line $k_y a=0$ cannot be explained by
CN. The crossing point between two branches of the helical edge mode
is protected by S symmetry, thus the 2D plane at $k_y a=0$ can be viewed as an AFMTI phase.
In addition, since the crossing point at ${\bf k}=0$ also falls into the mirror (or glide) plane, as shown
by diagonal line in Fig.\ref{WeylSE}c, that crossing point thus can also be protected by MCN, which
is equal to 1 for the phase II.
Therefore, two blue lines in Fig.\ref{WeylPD}a can be viewed
as transition lines between the phase II with $MCN=1$ and the phase I or III
with MCN being 0 or 2 respectively.

Although we focus on the magnetic moments parallel or perpendicular to the mirror plane in this section,
the WSM phase can NOT be destroyed immediately when magnetic moments is tilted away from
these directions due to the non-zero CNs carried by Weyl points.
The Weyl points can only move in the momentum space and should be robust in certain parameter regimes.
On the other hand, for the TMI phase, gapless points of helical surface mode at finite non-zero momenta are solely protected by the mirror or glide symmetry and thus
sensitively depend on the direction of magnetic moments.

\subsection{Nodal Line Semimetal Phase}
\begin{figure}[t]
\includegraphics[width=\columnwidth]{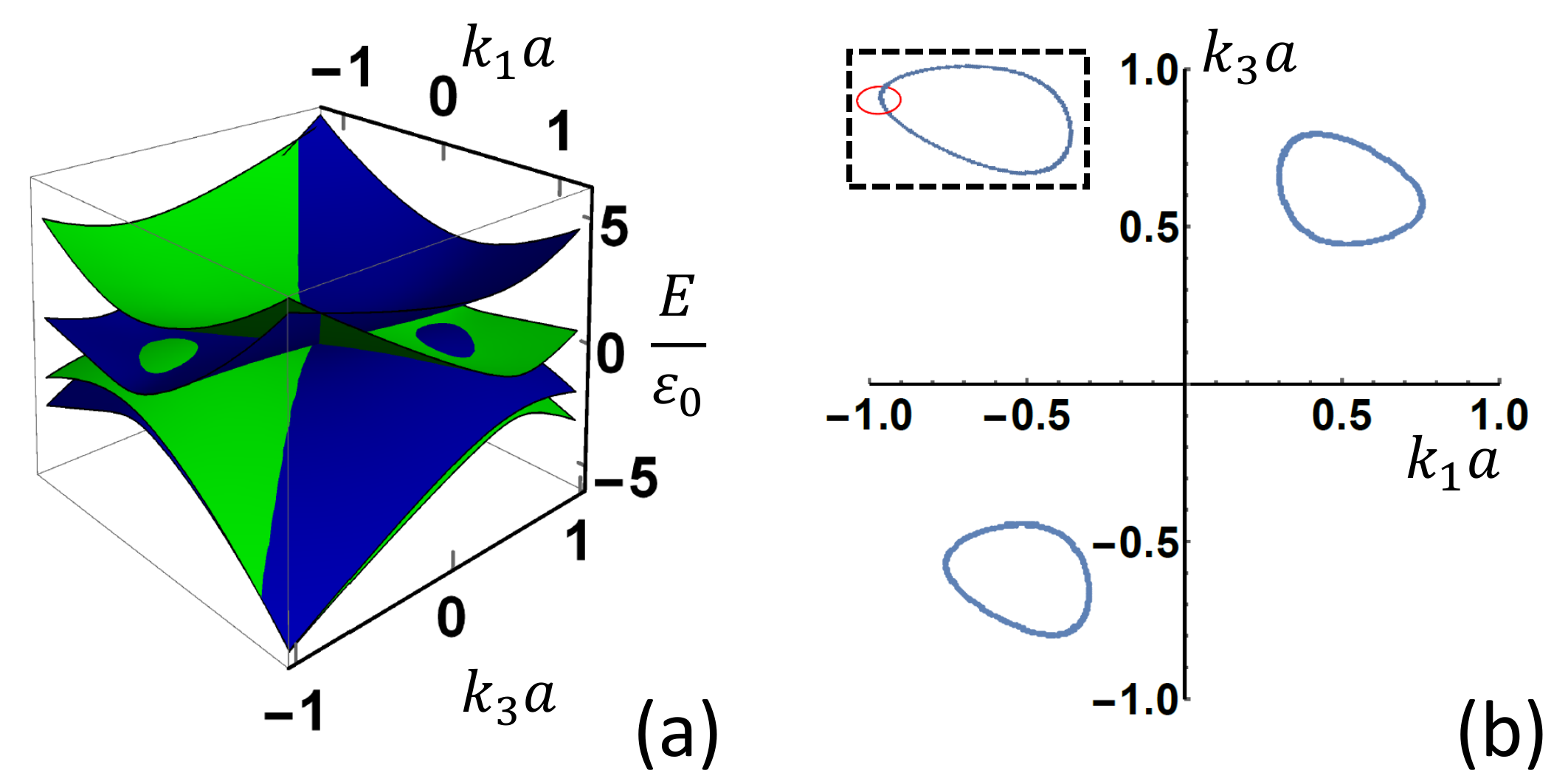}
\caption{
\label{Nodal}
(a)Energy dispersion on $(1\bar{1}0)$ plane
with green and blue bands coming from $+i$ and $-i$ mirror subspace, respectively.
Here $k_{1,2,3}$ are defined as $
(k_x,k_y,k_z)=
(\frac{k_1}{\sqrt{6}}-\frac{k_2}{\sqrt{2}}+\frac{k_3}{\sqrt{3}},\frac{k_1}{\sqrt{6}}+\frac{k_2}{\sqrt{2}}+\frac{k_3}{\sqrt{3}},-\frac{2 k_1}{\sqrt{6}}+\frac{k_3}{\sqrt{3}})$.
(b)The positions of nodal rings on $(1\bar{1}0)$ plane. The inset shows that
the red circle around the nodal ring , along which accumulated Berry phase is $\pi$.}
\end{figure}
For magnetic moments of anti-ferromagnetic ordering within or perpendicular to the $(1\bar{1}0)$ plane,
another possible TSM phase is NLSM phase.
An example of NLSM phase is shown in Fig.\ref{Nodal}, where
the parameters are shown in Tab.\ref{tab:Nodalfig} in Appendix \ref{app:parameters}.
Due to differnet AFM parameters, the band sequence is different and
two crossing bands at the low energy are from two opposite mirror subspaces,
in contrast to the band crossing between two bands with the same mirror parity
for the phase boundary in Fig.\ref{WeylPD}a. The energy dispersion of NLSM phase is
depicted in Fig.\ref{Nodal}a and the positions of two nodal rings are shown in Fig.\ref{Nodal}b.
The topological stability of each nodal ring can be extracted by the $\pi$ Berry phase
along a small circle (red circle in inset of Fig.\ref{Nodal}b) around the nodal line.

\subsection{Triple Point Semimetal Phase}
\label{sec:TP}
In this section, we consider the case with
anti-ferromagnetic magnetic moments along the $(111)$ direction,
where the system has three-fold rotational symmetry $C_3(111)$ and glide symmetry
$T_{\mathbf{a}'_3}\pi_{(1\bar{1}0)}$.
Since the matrix representation of $T_{\mathbf{a}'_3}\pi_{(1\bar{1}0)}$ is equivalent to
the mirror symmetry $\pi_{(1\bar{1}0)}$ for the basis of $\Gamma_8$ bands,
the symmetry group generated by $C_3(111)$ and $T_{\mathbf{a}'_3}\pi_{(1\bar{1}0)}$ is
isomorphic to the point group $C_{3v}$.
By linearly combining the four basis functions of $\Gamma_8$ bands which carry total angular momentum $J=\frac{3}{2}$,
we can get a pair of states belonging to
two-dimensional representation $\Lambda_6$ of the $C_{3v}$ double group, and the other two
states belonging to two one-dimensional representations $\Lambda_4$
and $\Lambda_5$ respectively.\cite{burns2014introduction} The character table of the $C_{3v}$ double group and the linear combinations of $\Gamma_8$ bases
are shown in Appendix \ref{app:c3v_double_group}.

\begin{figure}[t]
\includegraphics[width=\columnwidth]{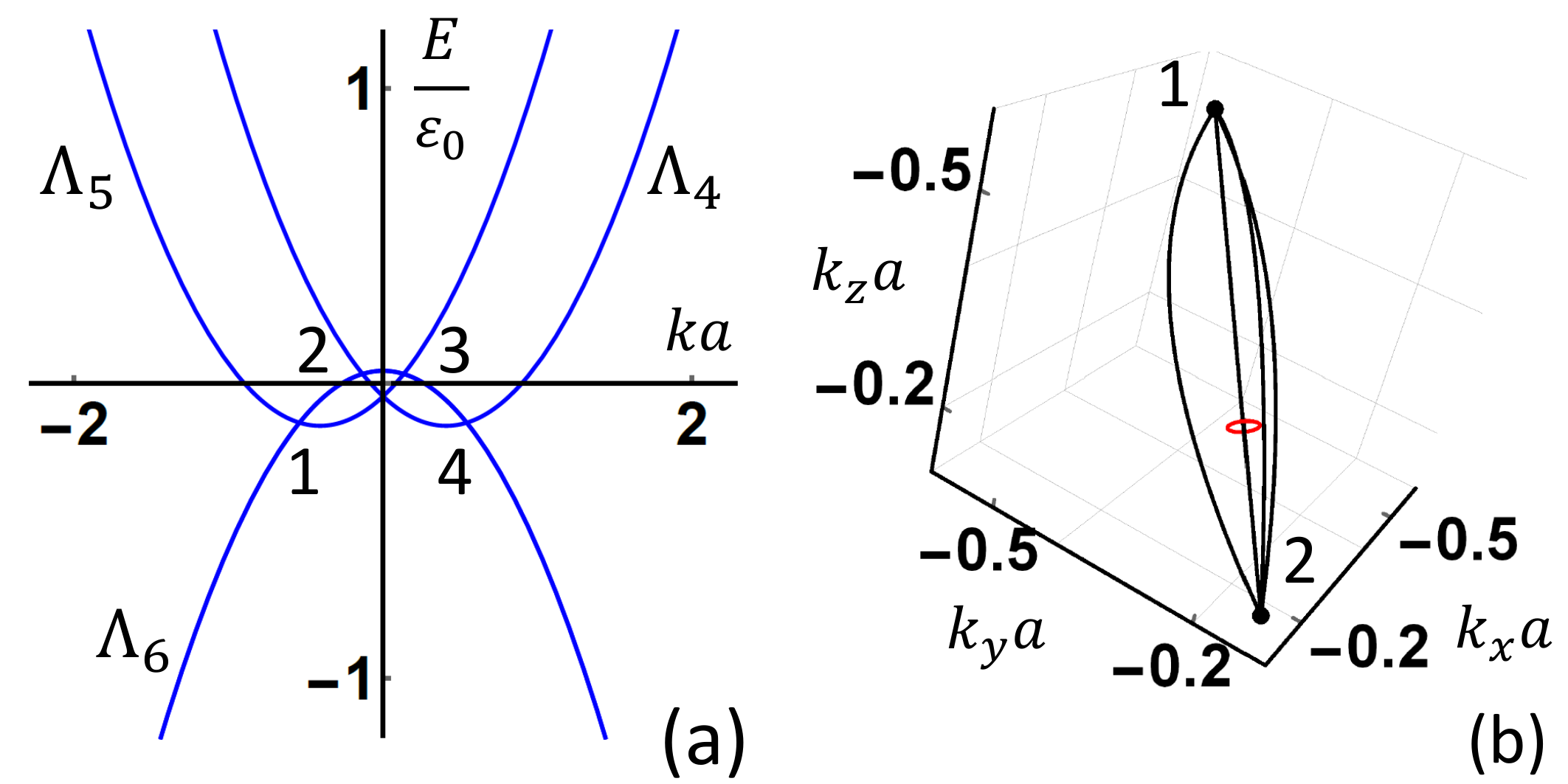}
\caption{
\label{TPSP}
(a)Band dispersion along the (111) direction for TPSM phase.
$\Lambda_4$ and $\Lambda_5$ label two different non-degenerate bands
whereas $\Lambda_6$ labels a doubly degenerate band.
The crossing points 1,2,3,4 are triple points with three-fold degeneracy.
(b) Black dots denote the positions for the points 1 and 2 in the momentum space and are
conntected by four nodal lines (black lines). Berry phase accumulated along the red circle around any nodal line
is $\pi$. }
\end{figure}

To confirm this symmetry analysis, we consider the total Hamiltonian
$H_0(\mathbf{k})+H_C(\mathbf{k})+H_{AFM}$ with magnetic moments along the $(111)$ direction
($M_x=M_y=M_z$). This corresponds to the conditions $\xi_1=\xi_2=0$ and $\xi_3=\xi_4=\xi_5$.
Along the $(111)$ direction, we indeed find that four $\Gamma_8$ bands are split into one doubly degenerate
band, labeled as $\Lambda_6$ bands, and another two non-degenerate bands, labeled as
 $\Lambda_4$ and $\Lambda_5$ bands respectively.
The corresponding energy dispersion can be solved analytically as
$$
\begin{array}{l}
E_{\Lambda_6}(k)=E_v+\xi_0-3\beta_c\gamma_1 k^2-(6 \beta_c \gamma_3 k^2 +\sqrt{3} \xi)\\
E_{\Lambda_5}(k)=E_v+\xi_0-3\beta_c\gamma_1 k^2+(6 \beta_c \gamma_3 k^2 +\sqrt{3} \xi)+\sqrt{6} C k\\
E_{\Lambda_4}(k)=E_v+\xi_0-3\beta_c\gamma_1 k^2+(6 \beta_c \gamma_3 k^2 +\sqrt{3} \xi)-\sqrt{6} C k
\end{array},
$$
where $\xi_3=\xi_4=\xi_5\equiv \xi$ and $k_x=k_y=k_z=k$.
A typical energy dispersion is shown in Fig.\ref{TPSP}a with the parameters shown in Tab.\ref{tab:TPfig} in Appendix \ref{app:parameters}.
Under the condition $C^2 >16 \sqrt{3} \beta_c \gamma_3 \xi$ and $C\xi\neq 0$,
the $\Lambda_6$ bands cross with the $\Lambda_5$ ($\Lambda_4$) band at two points
labeled by 1 and 3 (2 and 4) in Fig.\ref{TPSP}a.
At each crossing point, there is a three-fold degeneracy and the energy dispersion
behave linearly along the $(111)$ axis.
Points 1 and 2 (or 3 and 4) are connected by four nodal lines,
as shown in Fig.\ref{TPSP}b. Along a circle enclosing any of these four nodal lines, the accumulated Berry phase
is found to be $\pi$. This type of semi-metal phase is known as type-B TPSM phase \cite{PhysRevX.6.031003}.

\section{Anti-ferromagnetic Topological Insulating Phase in Six-band Kane Model}

\begin{figure}[t]
\includegraphics[width=\columnwidth]{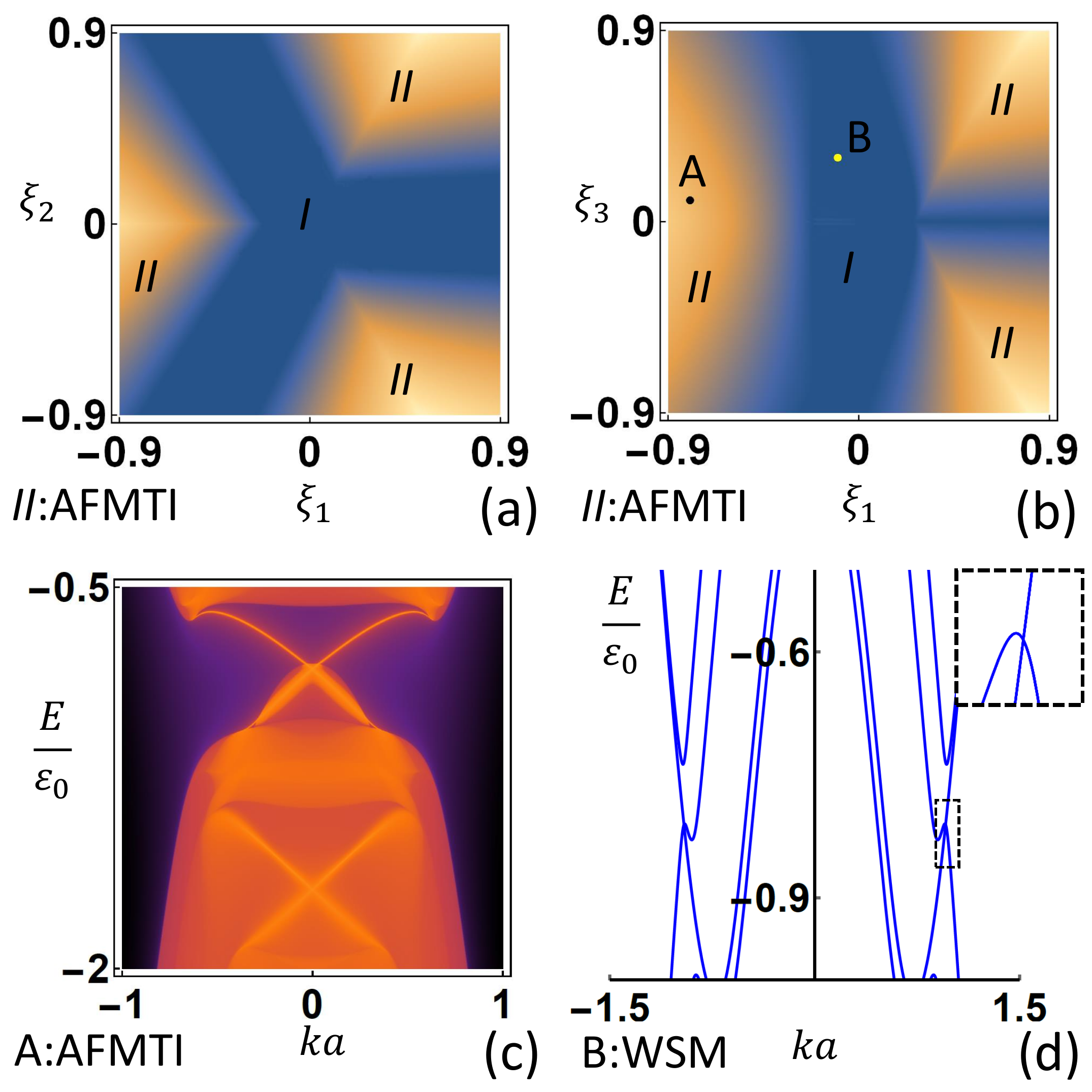}
\caption{
\label{6BPD0}
The direct band gap is shown as a function of (a) $\xi_1$ and $\xi_2$ for zero $\xi_{3,4,5}$
and (b) $\xi_1$ and $\xi_3$ for zero $\xi_{2,4,5}$.
Here the region I with the blue color is the TSM phase while the region II with the yellow color
has non-zero direct band gap and is the AFMTI phase.
(c) shows the local density of states on (001) surface for point A with $(\xi_1,\xi_3)=(-0.8,0.1)\varepsilon_0$ in (b). $k$ is along (110) direction.
(d) shows the bulk dispersion along the line connecting two Weyl points at $\mathbf{k}a=\pm(-0.89, -0.35, 0)$.
This dispersion corresponds to the point B in Fig.\ref{6BPD0}b with $(\xi_1,\xi_3)=(-0.1,0.3)\varepsilon_0$.
The inset is a zoom-in plot around the momentum $\mathbf{k}a=(-0.89, -0.35, 0)$.
}
\end{figure}

In the discussion above, we neglected $\Gamma_6$ bands and only focused on four $\Gamma_8$ bands.
This simplification has irrelevant influence on TSM phases since $\Gamma_6$ bands are far away
from Fermi energy.
However, due to the inverted nature between $\Gamma_6$ and $\Gamma_8$ bands, $\Gamma_6$ bands may
play an essential role for topological insulating phases.
It is well known that the band inversion between the
$\Gamma_6$ and $\Gamma_8$ bands leads to the topological insulating phase in HgTe \cite{Bernevig1757},
as well as non-magnetic half-Heusler materials
\cite{lin2010half,chadov2010tunable,PhysRevLett.105.096404,PhysRevB.82.125208}.
Therefore,we study the influence of $\Gamma_6$ bands by considering the full six-band Kane model $H_{Kane}$ in this section.
With the anti-ferromagnetic ordering, the full Hamitonian takes the form
\begin{equation}
\label{eq:h6}
H_{Kane}^{AFM}=
\left(
\begin{array}{c|c}
H_{\Gamma_6}(\mathbf{k})&V\\
\hline
V^{\dagger}&H_{\Gamma_8}(\mathbf{k})+H_{AFM}\\
\end{array}
\right),
\end{equation}
which can be diagonalized numerically to extract the energy dispersion.

To systematically understand the AFM term in the Kane model, we consider different $\xi_i$ terms ($i=1,\dots,5$)
in Eq.\ref{eq:Ham_AFM}, separately.
We notice that applying three-fold rotational symmetry $C_3(111)$ to $H_{AFM}$ is equivalent to
performing the following transformations: $(\xi_1,\xi_2)\rightarrow (-\frac{1}{2}\xi_1-\frac{\sqrt{3}}{2} \xi_2,\frac{\sqrt{3}}{2}\xi_1-\frac{1}{2}\xi_2)$ and $(\xi_3,\xi_4,\xi_5)\rightarrow (\xi_4,\xi_5,\xi_3)$, which means $\xi_{1,2}$ are related with each other by $C_3(111)$ and so do $\xi_{3,4,5}$.
Therefore, we study the direct band gap as a function of (i) $\xi_1$ and $\xi_2$ for $\xi_{3,4,5}=0$
in which case the effective Hamiltonian $H_{Kane}^{AFM}$ in Eq.\ref{eq:h6} preserves two-fold rotation symmetry along the x,y,z axes,
or (ii) $\xi_1$ and $\xi_3$ for $\xi_{2,4,5}=0$ in which case the effective model $H_{Kane}^{AFM}$ preserves $\pi_{1\bar{1}0}$,
two-fold rotation along z axis and mirror symmetry perpendicular to (110).
The phase diagrams for the case (i) and (ii) are shown in Fig.\ref{6BPD0}a and Fig.\ref{6BPD0}b, respectively,
from which one can find both the gapless phases existing in the blue region (the region I)
and the gaped phases in the three yellow regions (the region II).
Detailed parameters for our calculation can be found in Tab.\ref{tab:6BPD0(a)} and Tab.\ref{tab:6BPD0(b)} in Appendix \ref{app:parameters}.
We notice that $\xi_1$ term takes the same form as the strain term described in Ref.[\citen{ruan2016symmetry}].
Therefore, on the line $\xi_2=0$ in Fig.\ref{6BPD0}a or the line $\xi_3=0$ in Fig.\ref{6BPD0}b,
we expect the gapless and gaped phases should be equivalent to the corresponding ones studied
for strained HgTe and half-Heusler materials \cite{ruan2016symmetry}.
To verify the nature of gapless and insulating phases, we calculate the energy dispersion
for two typical sets of parameters: the point A with $(\xi_1,\xi_3)=(-0.8,0.1)\varepsilon_0$
and B with $(\xi_1,\xi_3)=(-0.1,0.3)\varepsilon_0$ in Fig.\ref{6BPD0}b. A non-zero bulk direct gap
is found for the point A and thus we consider an approximately semi-infinite configuration and plot the
local density of states on (001) surface, as shown in Fig.\ref{6BPD0}c.
A helical surface mode is found in the bulk gap and is protected by the $S$ symmetry
instead of the time reversal symmetry due to the anti-ferromagnetic ordering, thus giving
rise to a realization of AFMTI phase.
The gapless phase at point B is found to be WSM phase
and the bulk dispersion is shown Fig.\ref{6BPD0}d with the Weyl points located
at the momenta $\mathbf{k}a=\pm(0.89, 0.35, 0)$ (or equivalent $\mathbf{k}a=\pm(-0.35, -0.89, 0)$).

\begin{figure}[t]
\includegraphics[width=\columnwidth]{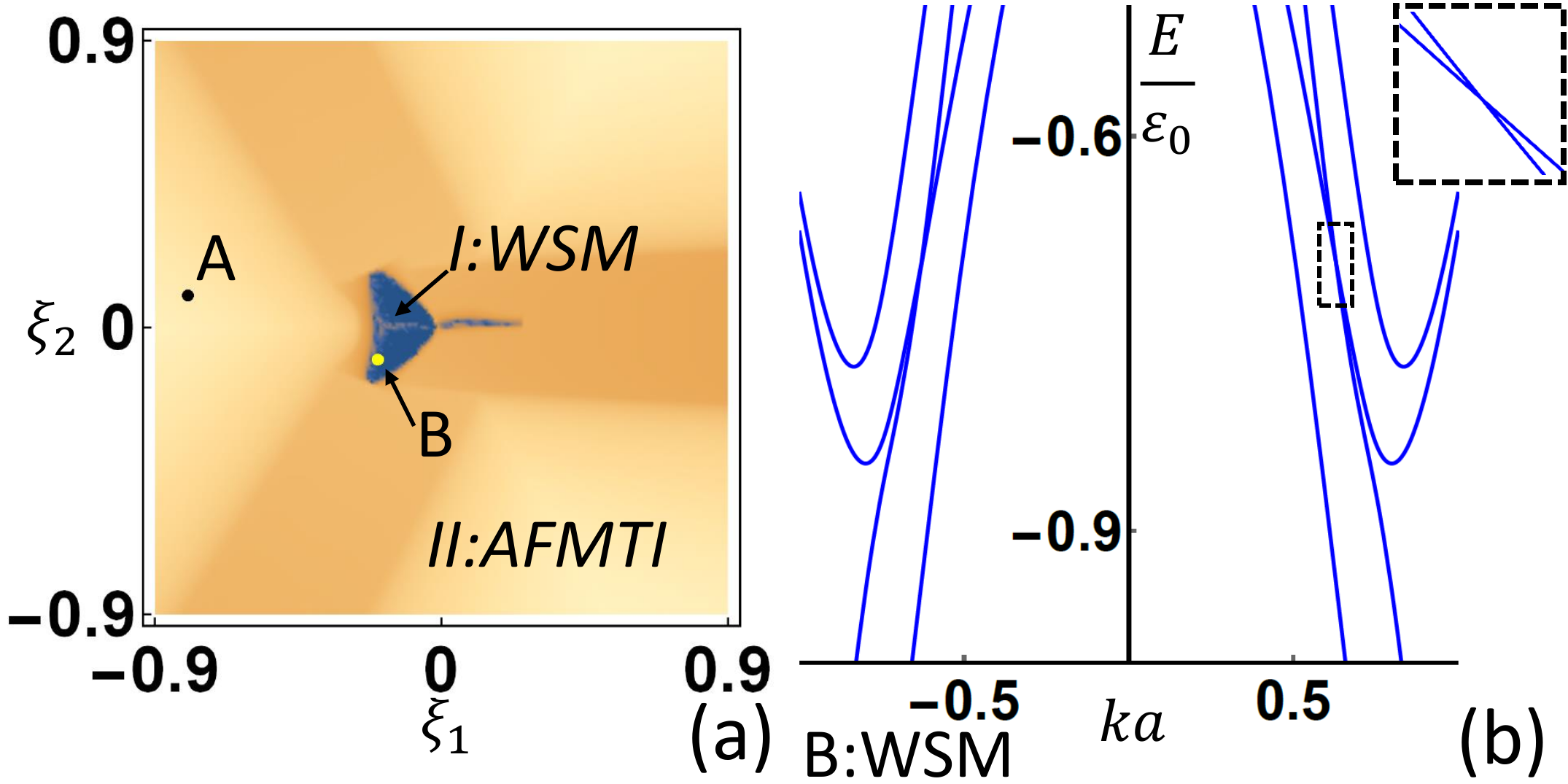}
\caption{
\label{6BPDF}
(a)  shows the approximate direct gap distribution of the system on $\xi_1-\xi_2$ parameter plane with non-zero $\xi_{3,4,5}$. The color is scaled by the logarithm with base 10 of the direct gap of system.
Lighter color indicates larger gap, and gaps smaller than $10^{-5}$ are marked with the same color as that of $10^{-5}$.
The graph shows two regions: (i) region I which should be WSM phase, and (ii) region II which should be AFMTI phase.
(b) shows the bulk dispersion along the line connecting two Weyl points at $\mathbf{k}a=\pm(-0.23, -0.21, 0.55)$. The Weyl points belong to point B in Fig.\ref{6BPDF}a with $(\xi_1,\xi_2)=(-0.2,-0.1)\varepsilon_0$.
The inset is a zoom-in version for $\mathbf{k}a=(-0.23, -0.21, 0.55)$.}
\end{figure}

Below we emphasize that the AFMTI phase is quite robust in this system.
The phase diagram for $\xi_1-\xi_2$ is shown in Fig.\ref{6BPDF}a for $(\xi_3,\xi_4,\xi_5)=(-0.01,0.023,0.026)\varepsilon_0$.
Other parameters are listed in Tab.\ref{tab:6BPDF} in Appendix \ref{app:parameters}. We notice that
after introducing non-zero $\xi_{3,4,5}$ which breaks all $C_{3v}$ symmetries and their combinations with half translation,
the previous gapless phase region I in Fig.\ref{6BPD0}a shrinks to a smaller region I
in Fig.\ref{6BPDF}a, while the region II of AFMTI phase is greatly extended.
In the region I, Weyl points are found (not exclusively) at $\mathbf{k}a=\pm(-0.23, -0.21, 0.55)$ for point B of Fig.\ref{6BPDF}a, as shown
in Fig.\ref{6BPDF}b. The local density of states calculation on (001) surface for point A in Fig.\ref{6BPDF}a
with $(\xi_1,\xi_2)=(-0.8,0.1)\varepsilon_0$ gives very similar graph as Fig.\ref{6BPD0}c,
thus demonstrating the AFMTI phase in the region II.
Given the large region in the parameter space for the realization of AFMTI phases,
we can conclude that anti-ferromagnetic half-Heusler materials provide
a robust material realization of the AFMTI phase.
Moreover, if mirror or non-symmorphic symmetry exists,
anti-ferromagnetic half-Heusler materials can also provide
a robust material realization of anti-ferromagnetic mirror or non-symmorphic topological insulator phase,
which has not been demonstrated in experiments.

\section{Conclusion}
\begin{figure}[t]
\includegraphics[width=\columnwidth]{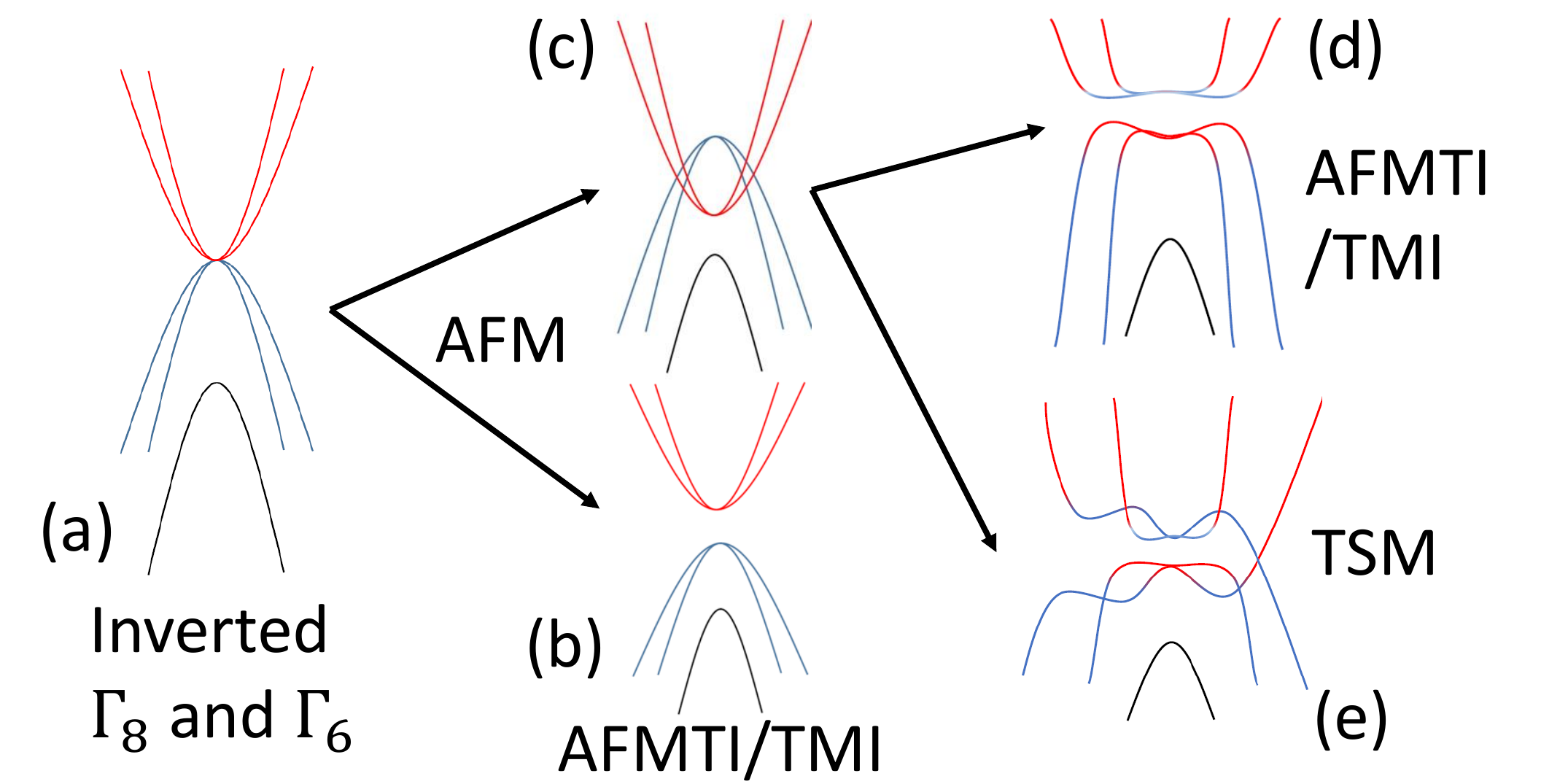}
\caption{
\label{fig:SchPhy}
Schematic physical picture of evolution of electronic band structures of half-Heusler compounds with AFM.
(a) is for the band ordering without AFM, while (b) and (c) are normal and inverted band ordering between different
$\Gamma_8$ bands after including AFM. Depending on detailed parameters for AFM, the inverted band structures
due to AFM can either be gapped or be gapless, leading to (d) topological insulating phases or (e) TSM phases.
}
\end{figure}

Based on the above studies of the four-band and six-band Kane models, we can summarize
the overall physical picture for electronic structures of anti-ferromagnetic half-Heusler compounds in Fig.\ref{fig:SchPhy}.
Without AFM, the band ordering of $\Gamma_6$ and $\Gamma_8$ bands are inverted and the Fermi energy
lies at the four fold degenerate point ($\mathbf{k}=0$) of $\Gamma_8$ bands, leading to a critical semimetal phase shown in Fig.\ref{fig:SchPhy}a.
With AFM, four-fold degeneracy of $\Gamma_8$ bands at $\Gamma$ point is removed. The band structure of four $\Gamma_8$
bands can be either normal (Fig.\ref{fig:SchPhy}b) or inverted (Fig.\ref{fig:SchPhy}c), depending on
the detailed form and parameters of AFM terms.
When the band ordering of $\Gamma_8$ bands is normal, it is trivial for the four-band model but non-trivial for the six-band Kane model due to the inversion between the $\Gamma_6$ and $\Gamma_8$
bands, leading to either AFMTI phase or TMI phase. When the $\Gamma_8$ band ordering is inverted,
the AFM terms can either lead to a full inverted band gap (AFMTI or TMI phase in Fig.\ref{fig:SchPhy}d)
or preserve certain gapless points
in the momentum space (the TSM phase in Fig.\ref{fig:SchPhy}e).
In either situation, we find that anti-ferromagnetic half-Heusler compounds are
topologically non-trivial.
Thus, our work demonstrates that half-Heusler materials with AFM provide a platform
for a robust realization of anti-ferromagnetic topological phases, either WSM phase or AFMTI phase, in a wide parameter regime.

The AFMTI phase was first proposed in Ref.[\citen{moore2010}] based on a four band toy model and our results have shown
this interesting topological phase indeed can exist in anti-ferromagnetic half-Heusler materials.
We notice that the first principles calculation in combining with tight-binding model has been adopted for the
AFM GdPtBi\cite{li2015electronic}, in which a semi-metal phase is found.
However, the topological nature of this semi-metal phase
has not been extracted and our results identify the existence of Weyl points in this semimetal phase.
In addition, the authors use the representation of crystal symmetry group to label each band,
aiming in identifying band inversion.
We believe this approach is insufficient for the AFMTI phase
since this topological phase is protected by the $S$ symmetry which is not included
in the crystal symmetry group. Thus, the AFMTI phase cannot be identified from the crystal symmetry representation.
Our results suggest that the AFMTI phase can exist in the G-type anti-ferromagnetic system, which is identified to be
topologically trivial in Ref.[\citen{li2015electronic}].
In the existing experiments, Weyl semi-metal phase has been unveiled in GdPtBi under an external
magentic field through the observation of a large anomalous Hall angle \cite{suzuki2016large,shekhar2016observation}, large negative magnetoresistance\cite{shekhar2016observation,Hirschberger2016}
and the strong suppression of thermopower\cite{Hirschberger2016}.
Our results suggest Weyl semi-metal phase may already occur even in absence of external magnetic fields.
In addition, nodal line fermions, type-B triply degenerate fermions and
topological mirror or glide insulators, are also possible in certain parameter regimes when
magnetic moments of anti-ferromagnetic ordering are along some specific directions.
The topological surface states or surface Fermi arcs in these topological phases can be extracted
from angle-resolved photoemission spectroscopy or quasi-particle spectrum from
the scanning tunneling microscopy \cite{RevModPhys.83.1057,RevModPhys.82.3045,yan2016topological}.
Our generalized Kane model also provides a basis for the future study of magnetic,
transport or optical phenomena in this class of materials.
Furthermore, we notice that superconductivity can coexist with anti-ferromagnetism in
RPdBi(R=Tb,Ho,Dy,Er)\cite{nakajima2015topological,pan2013superconductivity,pavlosiuk2016antiferromagnetism}.
Thus, it is interesting to ask if topological superconductivity can be realized in these materials.

\section{Acknowledgment}
We would like to thank Rui-Xing Zhang, Jian-Xiao Zhang, Qing-Ze Wang, Yang Ge and Di Xiao for helpful discussion. C.-X.L. acknowledges the support from Office of Naval Research (Grant No. N00014-15-1-2675).
B.Y. acknowledges support of the Ruth and Herman Albert Scholars Program for New Scientists in Weizmann Institute of Science, Israel.

%

\appendix

\section{Tight-Binding Model and its relationship to the Kane model}
\label{app:Tight-binding}
In this section, we will describe a tight-binding model for our anti-ferromagnetic half-Heusler materials, from which we can
justify the extend Kane model that we derived by symmetry principles and used for the low energy physics in Sec.\ref{sec:Model Hamiltonian of anti-ferromagnetic half-Heusler materials}
in the main text.
For simplicity, we take
the anti-ferromagnetic half-Heusler material ErPdBi as an example and assume it has anti-ferromagnetic structure of GdPtBi. We consider nine orbitals
$\left|Bi,s\right\rangle$, $\left|Pd, s\right\rangle$, $\left|Pd,p_x\right\rangle$, $\left|Pd,p_y\right\rangle$, $\left|Pd,p_z\right\rangle$, $\left|Er,s\right\rangle$, $\left|Er,d_{xy}\right\rangle$, $\left|Er,d_{yz}\right\rangle$ and $\left|Er,d_{zx}\right\rangle$ to construct the tight-binding model for this material.
Due to the doubling of the unit cell along $\mathbf{a}'_3$,
all the orbitals are labeled as $\left|n, i\right\rangle$, where $n$ stands for atoms and orbitals and
$i=1,2$ labels two sub-lattices that are related by $T_{\mathbf{a}'_3}$.
We only considered the hopping terms between the nearest neighbor atoms, including Bi, Pd and Er atoms \cite{PhysRev.94.1498}, giving rise to a 18 by 18 Hamiltonian $H_{hopping}$.
In order to include AFM term and spin-orbit term, we need to consider spin degree of freedom and the hopping term
is block diagonal with each block given by $H_{hopping}$ in the spin space.
The AFM term is given by $H_{AFM}=(-1)^i\mathbf{M}\cdot\mathbf{S}$ with $i=1,2$ for two sets of cells.
Here $\mathbf{M}$ is the mean field value for anti-ferromagnetic magnetic moments and $\mathbf{S}$ labels electron spin.
The spin-orbit coupling term takes the form $H_{so}=\mathbf{S}\cdot\mathbf{L}$, where $\mathbf{L}$ denotes
the angular momentum operator that acts on the oribtal ($s, p, d$ orbitals).

\begin{figure}[H]
\includegraphics[width=\columnwidth]{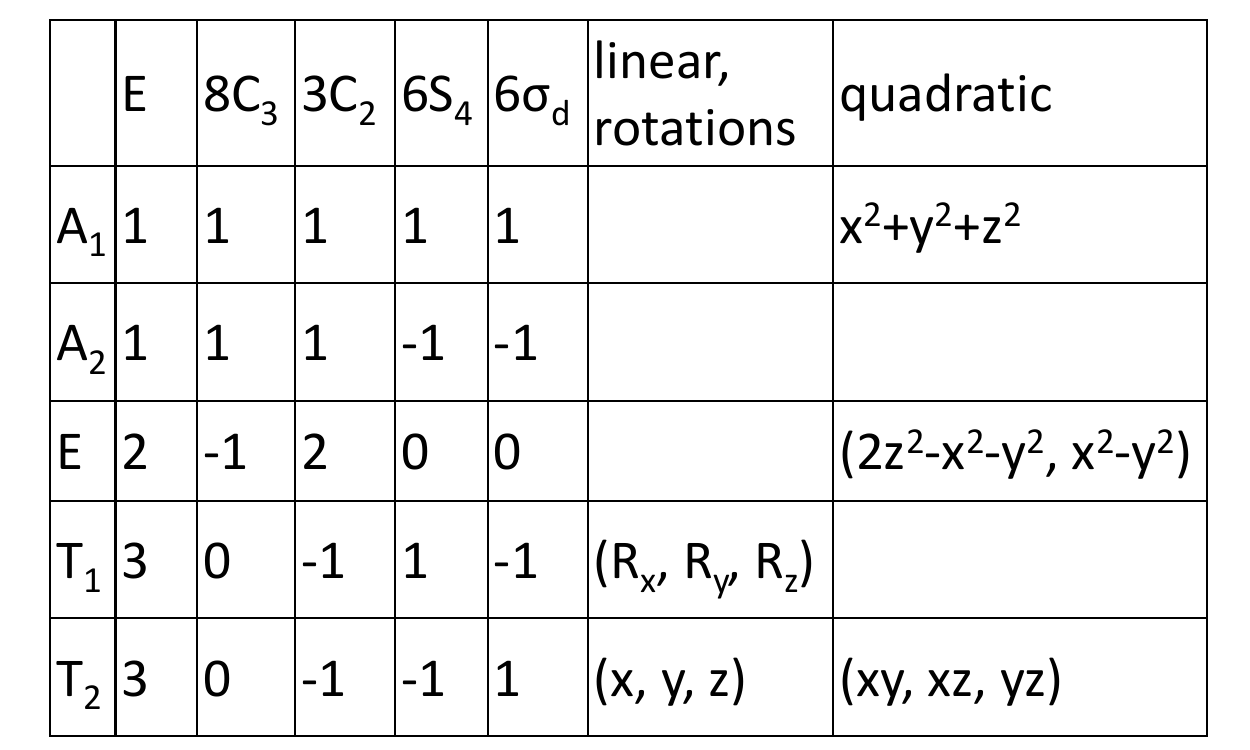}
\caption{Character Table of $T_d$ group\cite{gelessus1995multipoles,aroyo2006bilbao}. $p_x,p_y,p_z$ orbitals and $d_{xy},d_{yz},d_{zx}$ orbitals belong to $T_2$ representation.
\label{fig:Td}}
\end{figure}

The above 36 by 36 tight-binding Hamiltonian is complicated and we are not interested in using this model
for realistic calculation here. Instead, we hope to use this model to justify the form of the extended Kane model
that we derived from the symmetry principle in the main text.
We need to transform the basis wave functions into the ones used for the Kane model, which can be achieved by
the following steps. (1) The basis wave functions of the Kane model only include two $s$ orbital type of bands and 6
$p$ orbital type of bands after taking into account spin degree of freedom. Therefore, we need to first reduce
the number of basis wave functions.
Since both the $p_x,p_y,p_z$ and $d_{xy},d_{yz},d_{zx}$ orbitals belong to $T_2$ representation
of the $T_d$ group (The representation table of $T_d$ group is shown in Fig.\ref{fig:Td}),
the basis wave functions should be a linear combination of the $p$ and $d$ orbitals,
which are given by $\left|\gamma,i,\sigma\right\rangle=P_1\left|Pd,p_{\alpha},i,\sigma\right\rangle+P_2\left|Er,d_{\beta},i,\sigma\right\rangle$, where $i=1,2$, $\sigma=\uparrow,\downarrow$ and $(\gamma,\alpha,\beta)=(X,x,yz),(Y,y,zx),(Z,z,xy)$.
For the s orbital type of bands, the basis wave functions take the form
$\left|S,i,\sigma\right\rangle=S_1\left|Bi,s,i,\sigma\right\rangle+S_2\left|Pd,s,i,\sigma\right\rangle+S_3\left|Er,s,i,\sigma\right\rangle$, where $i=1,2$ and $\sigma=\uparrow,\downarrow$.
Here we still keep the sub-lattice index $i=1,2$ and the number of basis wave functions are reduced to 16.
(2) The basis wave functions of the Kane model is chosen to be the eigenstates of total angular momentum operator
because of strong spin-orbit coupling. Therefore, we apply a similar transformation to our basis wave functions here,
as listed below:
$$
\begin{array}{c}
\left|\Gamma_6,\frac{1}{2},1\right\rangle=i \left|S,\uparrow,1\right\rangle\\
\left|\Gamma_6,-\frac{1}{2},1\right\rangle=i \left|S,\downarrow,1\right\rangle\\
\left|\Gamma_8,\frac{3}{2},1\right\rangle=-\frac{1}{\sqrt{2}}\left|X,\uparrow,1\right\rangle+\frac{i}{\sqrt{2}}\left|Y,\uparrow,1\right\rangle\\
\left|\Gamma_8,\frac{1}{2},1\right\rangle=-\frac{1}{\sqrt{6}}\left|X,\downarrow,1\right\rangle+\frac{i}{\sqrt{6}}\left|Y,\downarrow,1\right\rangle+\sqrt{\frac{2}{3}}\left|Z,\uparrow,1\right\rangle\\
\left|\Gamma_8,-\frac{1}{2},1\right\rangle=\frac{1}{\sqrt{6}}\left|X,\uparrow,1\right\rangle+\frac{i}{\sqrt{6}}\left|Y,\uparrow,1\right\rangle+\sqrt{\frac{2}{3}}\left|Z,\downarrow,1\right\rangle\\
\left|\Gamma_8,-\frac{3}{2},1\right\rangle=\frac{1}{\sqrt{2}}\left|X,\downarrow,1\right\rangle+\frac{i}{\sqrt{2}}\left|Y,\downarrow,1\right\rangle\\
\left|\Gamma_7,\frac{1}{2},1\right\rangle=-\frac{1}{\sqrt{3}}\left|X,\downarrow,1\right\rangle+\frac{i}{\sqrt{3}}\left|Y,\downarrow,1\right\rangle-\frac{1}{\sqrt{3}}\left|Z,\uparrow,1\right\rangle\\
\left|\Gamma_7,-\frac{1}{2},1\right\rangle=-\frac{1}{\sqrt{3}}\left|X,\uparrow,1\right\rangle-\frac{i}{\sqrt{3}}\left|Y,\uparrow,1\right\rangle+\frac{1}{\sqrt{3}}\left|Z,\downarrow,1\right\rangle\\
\left|\Gamma_6,\frac{1}{2},2\right\rangle=i \left|S,\uparrow,2\right\rangle\\
\left|\Gamma_6,-\frac{1}{2},2\right\rangle=i \left|S,\downarrow,2\right\rangle\\
\left|\Gamma_8,\frac{3}{2},2\right\rangle=-\frac{1}{\sqrt{2}}\left|X,\uparrow,2\right\rangle+\frac{i}{\sqrt{2}}\left|Y,\uparrow,2\right\rangle\\
\left|\Gamma_8,\frac{1}{2},2\right\rangle=-\frac{1}{\sqrt{6}}\left|X,\downarrow,2\right\rangle+\frac{i}{\sqrt{6}}\left|Y,\downarrow,2\right\rangle+\sqrt{\frac{2}{3}}\left|Z,\uparrow,2\right\rangle\\
\left|\Gamma_8,-\frac{1}{2},2\right\rangle=\frac{1}{\sqrt{6}}\left|X,\uparrow,2\right\rangle+\frac{i}{\sqrt{6}}\left|Y,\uparrow,2\right\rangle+\sqrt{\frac{2}{3}}\left|Z,\downarrow,2\right\rangle\\
\left|\Gamma_8,-\frac{3}{2},2\right\rangle=\frac{1}{\sqrt{2}}\left|X,\downarrow,2\right\rangle+\frac{i}{\sqrt{2}}\left|Y,\downarrow,2\right\rangle\\
\left|\Gamma_7,\frac{1}{2},2\right\rangle=-\frac{1}{\sqrt{3}}\left|X,\downarrow,2\right\rangle+\frac{i}{\sqrt{3}}\left|Y,\downarrow,2\right\rangle-\frac{1}{\sqrt{3}}\left|Z,\uparrow,2\right\rangle\\
\left|\Gamma_7,-\frac{1}{2},2\right\rangle=-\frac{1}{\sqrt{3}}\left|X,\uparrow,2\right\rangle-\frac{i}{\sqrt{3}}\left|Y,\uparrow,2\right\rangle+\frac{1}{\sqrt{3}}\left|Z,\downarrow,2\right\rangle\\
\end{array},
$$
where $1,2$ are sub-lattice indexes.
(3) Since the low energy physics only occurs at the $\Gamma$ point (${\bf k}=0$) for the Kane model,
we only need to consider the bonding states between two sub-lattices. The 8 bonding basis wave functions
are given by $|\Lambda,\sigma\rangle=\frac{1}{\sqrt{2}}(|\Lambda,\sigma,1\rangle+|\Lambda,\sigma,2\rangle)$,
where $\Lambda=\Gamma_{6,7,8}$ and $\sigma=\pm\frac{1}{2}$,or including $\pm\frac{3}{2}$ for $\Gamma_8$.
By projecting the full tight-binding Hamiltonian into the six basis wave functions, including
$|\Gamma_6,\frac{1}{2}\rangle$, $|\Gamma_6,-\frac{1}{2}\rangle$, $|\Gamma_8,\frac{3}{2}\rangle$,
$|\Gamma_8,\frac{1}{2}\rangle$, $|\Gamma_8,-\frac{1}{2}\rangle$ and $|\Gamma_8,-\frac{3}{2}\rangle$
and expand the resulting Hamiltonian to the second order in both ${\bf k}$ and ${\bf M}$,
we reproduce the extended Kane model Hamiltonian with the anti-ferromagnetic ordering derived by the symmetry principles
in our main text.

\section{Irreducible Representations of $C_{3v}$ group}
\label{app:irrep_C3v}
In this section, we will use the representation table of the $C_{3v}$ group to classify the polynomials
of ${\bf k}$ and ${\bf M}$, as well as all the four by four matrices for the $\Gamma_8$ basis wave functions,
which is used to construct the effective low energy Hamiltonian for our system
in Sec.\ref{sec:Model Hamiltonian of anti-ferromagnetic half-Heusler materials}.

The $C_{3v}$ group can be generated by two operations, three-fold rotation and mirror operation.
For the lattice considered here, the three-fold rotation is along the $(111)$ axis, denoted as $C_3(111)$,
and mirror operation is with repsect to the $(1,-1,0)$ plane, denoted as $\pi_{1\bar{1}0}$, where $\bar{1}$ means $-1$.
It has three irreducible representations $A_1$, $A_2$ and $E$ and its character table is shown in Tab.\ref{tab:C3v}.
\begin{table}
$$
\
\begin{array}{|c|c|c|c|}
\hline
&E&2C_3&3\sigma_v\\
\hline
A_1&	+1&	+1&	+1\\
\hline
A_2&	+1&	+1&	-1\\
\hline
E&	+2&	-1&	0\\
\hline
\end{array}
$$
\caption{\label{tab:C3v}
Character table of $C_{3v}$ group.\cite{gelessus1995multipoles,aroyo2006bilbao}}
\end{table}

Since magnetic moments of the anti-ferromagnetic ordering come from the $d$ orbitals of $Er$ atoms,
we only focus on the the $\Gamma_8$ bands which contains $d$ orbitals of $Er$ atoms.
We need to construct the generator operations of the $C_{3v}$ group on the $\Gamma_8$ bands.
To achieve that, we notice the angular momentum operators $J_{x,y,z}$ for spin-$\frac{3}{2}$ states
can be represented by the four by four matrices
$$
J_x\doteq\left(
\begin{array}{cccc}
 0 & \frac{\sqrt{3}}{2} & 0 & 0 \\
 \frac{\sqrt{3}}{2} & 0 & 1 & 0 \\
 0 & 1 & 0 & \frac{\sqrt{3}}{2} \\
 0 & 0 & \frac{\sqrt{3}}{2} & 0 \\
\end{array}
\right)
$$
$$
J_y\doteq\left(
\begin{array}{cccc}
 0 & -\frac{i \sqrt{3}}{2} & 0 & 0 \\
 \frac{i \sqrt{3}}{2} & 0 & -i & 0 \\
 0 & i & 0 & -\frac{i \sqrt{3}}{2}  \\
 0 & 0 & \frac{i \sqrt{3}}{2} & 0 \\
\end{array}
\right)
$$
$$
J_z\doteq\left(
\begin{array}{cccc}
 \frac{3}{2} & 0 & 0 & 0 \\
 0 & \frac{1}{2} & 0 & 0 \\
 0 & 0 & -\frac{1}{2} & 0 \\
 0 & 0 & 0 & -\frac{3}{2} \\
\end{array}
\right).
$$
As a result, the $C_3(111)$ and $\pi_{1\bar{1}0}$ operations are given by
$$
C_3(111)\doteq \exp(- i \frac{J_x+J_y+J_z}{\sqrt{3}} \frac{2 \pi}{3})
$$
$$
\pi_{1\bar{1}0}\doteq -\exp(- i \frac{J_x-J_y}{\sqrt{2}} \pi).
$$
In addition, time reversal operator and half translation operator $T_{\mathbf{a}'_3}$
are interesting to us since the anti-ferromagnetic ordering preserves the S symmetry which is the
combination of these two operators. On the four basis of the $\Gamma_8$ bands, we find
time reversal operator writeen as
$$
\Theta\doteq\left(
\begin{array}{cccc}
 0 & 0 & 0 & -1 \\
 0 & 0 & 1 & 0 \\
 0 & -1 & 0 & 0 \\
 1 & 0 & 0 & 0 \\
\end{array}
\right)\hat{K},
$$
with the complex conjuate $\hat{K}$ and the half translation $T_{\mathbf{a}'_3}$ taking the form
of an identity matrix
$$
T_{\mathbf{a}'_3}\doteq\left(
\begin{array}{cccc}
 1 & 0 & 0 & 0 \\
 0 & 1 & 0 & 0 \\
 0 & 0 & 1 & 0 \\
 0 & 0 & 0 & 1 \\
\end{array}
\right),
$$
because of the bonding state nature of two sub-lattices for the four $\Gamma_8$ bands.
As a consequence, the S symmetry takes the same form as the time reversal symmetry
on the four $\Gamma_8$ basis.

16 four-by-four matrices can be constructed by three spin $\frac{3}{2}$ matrices $J_{x,y,z}$
as  $J_x$, $J_y$, $J_z$, $J_x^2$, $J_y^2$, $J_z^2$, $J_{xy}=\{J_x,J_y\}/2$, $J_{yz}=\{J_y,J_z\}/2$, $J_{zx}=\{J_z,J_x\}/2$, $J_x^3$, $J_y^3$, $J_z^3$, $V_x=\{J_y^2-J_z^2,J_x\}/2$, $V_y=\{J_z^2-J_x^2,J_y\}/2$, $V_z=\{J_x^2-J_y^2,J_z\}/2$ and $J_{xyz}=J_xJ_yJ_z+J_zJ_yJ_x$.
The linear combinations of these 16 matrices belong to different irreducible representations of the $C_{3v}$ group and
we list the corresponding representations in in Tab.\ref{tab:C3vJ}.

\begin{table}[t]
$$
\begin{array}{|c|c|c|c|}
\hline
C_{3v}&$Combinations of $J_i$'s$& \Theta&T_{\mathbf{a}'_3}\\
\hline
A_1&J_x^2+J_y^2+J_z^2& + & +\\
\hline
A_1&J_{xy}+J_{yz}+J_{zx}& + & +\\
\hline
A_1&V_x+V_y+V_z& - & +\\
\Xhline{4\arrayrulewidth}
A_2& J_{xyz}& -& +\\
\hline
A_2& J_x+J_y+J_z& -& +\\
\hline
A_2& J_x^3+J_y^3+J_z^3& -& +\\
\Xhline{4\arrayrulewidth}
E&\sqrt{3}(-J_{x}+J_{y});\ 2 J_{z}-J_{x}-J_{y}& -& +\\
\hline
E&\sqrt{3}(-J_{x}^3+J_{y}^3);\ 2 J_{z}^3-J_{x}^3-J_{y}^3& -& +\\
\hline
E&2 V_{z}-V_{x}-V_{y};\ \sqrt{3}(V_{x}-V_{y})& -& +\\
\hline
E&2 J_{z}^2-J_{x}^2-J_{y}^2;\ \sqrt{3}(J_{x}^2-J_{y}^2)& +& +\\
\hline
E&2 J_{xy}-J_{yz}-J_{zx};\ \sqrt{3}(J_{yz}-J_{zx})& +& +\\
\hline
\end{array}
$$
\caption{
\label{tab:C3vJ}
This table shows combinations of $J_i$'s, their corresponding $C_{3v}$ irreducible representations and their parities under time reversal $\Theta$ and half translation $T_{\mathbf{a}'_3}$.
The first column shows the corresponding irreducible representations of $C_3v$.
The second column shows the expressions of combinations.
The third and fourth columns show how signs of those combinations change under $\Theta$ and $T_{\mathbf{a}'_3}$
, where ``$+$'' means no sign change and ``$-$'' means the sign should change.}
\end{table}

Next we will classify all the polynomials of ${\bf k}$ and ${\bf M}$ according to the irreducible
representations of the $C_{3v}$ group \cite{winkler2003spin}.
The representations for the polynomials of the momentum ${\bf k}$ and anti-ferromagnetic magnetic moments ${\bf M}$
are listed in Tab.\ref{tab:C3vkM}.

\begin{table}[t]
$$
\begin{array}{|c|c|c|c|}
\hline
C_{3v}&$Combinations of $k_i$'s or $M_i$'s$& \Theta&T_{\mathbf{a}'_3}\\
 \hline
A_1&k_x+k_y+k_z& -& +\\
\hline
A_1& k_x^2+k_y^2+k_z^2& +& +\\
\hline
A_1& k_xk_y+k_yk_z+k_zk_x& +& +\\
\hline
A_1& M_x^2+M_y^2+M_z^2& +& +\\
\hline
A_1& M_x M_y+M_y M_z+M_z M_x& +& +\\
\Xhline{4\arrayrulewidth}
A_2&M_x+M_y+M_z& -& -\\
\Xhline{4\arrayrulewidth}
E&2 k_{z}-k_{x}-k_{y};\ \sqrt{3}(k_{x}-k_{y})& -& +\\
\hline
E&2 k_{z}^2-k_{x}^2-k_{y}^2;\ \sqrt{3}(k_{x}^2-k_{y}^2)& +& +\\
\hline
E&2 k_x k_y-k_y k_z-k_z k_x;\ \sqrt{3}(k_y k_z-k_z k_x)& +& +\\
\hline
E&\sqrt{3}(-M_{x}+M_{y});\ 2 M_{z}-M_{x}-M_{y}& -& -\\
\hline
E&2 M_{z}^2-M_{x}^2-M_{y}^2;\ \sqrt{3}(M_{x}^2-M_{y}^2)& +& +\\
\hline
E&2 M_x M_y-(M_y+M_x) M_z;\ \sqrt{3}M_z(M_y -M_x)& +& +\\
\hline
\end{array}
$$
\caption{
\label{tab:C3vkM}
This table shows combinations of $k_i$'s or $M_i$'s, their corresponding $C_{3v}$ irreducible representations and their parities under time reversal $\Theta$ and half translation $T_{\mathbf{a}'_3}$.
The first column shows the corresponding irreducible representations of $C_{3v}$.
The second column shows the expressions of combinations.
The third and fourth columns show how signs of those combinations change under $\Theta$ and $T_{\mathbf{a}'_3}$, where ``$+$'' means no sign change and ``$-$'' means the sign should change.
}
\end{table}

\section{Gapless Conditions for the Hamiltonian $H_0+H_{AFM}$}
\label{app:gapless_conditions_4band0}
In this section, we will present the detailed analysis of the conditions for
possible gapless phases for the four-by-four Hamiltonian $H_0+H_{AFM}$, which
is discussed in Sec.\ref{sec:Dirac Semimetal Phase and Topological Mirror Insulating Phase} in the main text to understand the phase diagram
of the above Hamiltonian.
Without anti-ferromagnetic ordering, the energy dispersion of the Luttinger Hamiltonian $H_0$
possesses a quadratic touching at the $\Gamma$ point. With the AFM term
(at least one nonzero $\xi_i$ for $i\neq 0$), the Hamiltonian $H_0+H_{AFM}$ can be analytically
solved (Eq.\ref{eq:0+AFM}). The requirement for a gapless point is that $h_i({\bf k})+\xi_i=0$
for all $i=1,2,\dots,5$ for a certain momentum ${\bf k}$.
Below we will list all the possible cases for a gapless point to exist in the momentum
space. Here $\beta_c,\gamma_2,\gamma_3 \neq 0$ are always assumed.

{\bf Case I}: $\xi_3=\xi_4=\xi_5=0$.

In this case, we require $h_3=h_4=h_5=0$. According to the form of $h_{3,4,5}$, we immediately
see that two of three components of the momentum $(k_x,k_y,k_z)$ should be zero.
Let us assume $k_x=k_y=0$ and for the remaining non-zero $k_z$,
we still need to solve two equations $h_1=-\xi_1$ and $h_2=-\xi_2$.
This suggests that $\xi_1$ and $\xi_2$ cannot be independent of each other in order to achieve
a gapless phase.
According to the form of $h_{1,2}$, we find gapless points at
$$
\mathbf{k}a=(0,0,\pm \sqrt{-\frac{\xi_1}{2\beta_c\gamma_2}})
$$
for $\xi_3=\xi_4=\xi_5=0,\xi_2= 0,\xi_1 \beta_c \gamma_2<0$.

Similarly, if $k_y=k_z=0$, we have gapless points at
$$
\mathbf{k}a=\pm(\sqrt{-\frac{\xi_2}{\sqrt{3}\beta_c\gamma_2}},0,0)
$$
for $\xi_3=\xi_4=\xi_5=0,\xi_2\beta_c\gamma_2< 0,\xi_2=-\sqrt{3}\xi_1$.

And if $k_x=k_z=0$, we have gapless points at
$$
\mathbf{k}a=\pm(0,\sqrt{\frac{\xi_2}{\sqrt{3}\beta_c\gamma_2}},0)
$$
for $\xi_3=\xi_4=\xi_5=0,\xi_2\beta_c\gamma_2> 0,\xi_2=\sqrt{3}\xi_1$.

{\bf Case II}: For $\xi_3$, $\xi_4$ and $\xi_5$, two of them are zero and one is non-zero.

For this case, let's take the example of $\xi_3=\xi_4=0$ and $\xi_5\neq0$ and the analysis for
other cases are similar. Since $\xi_5\neq0$,
$k_{y,z}$ cannot be zero from $h_5=-\xi_5$. Therefore, $\xi_3=\xi_4=0$ gives rise to $k_x=0$.
For the remaining three equations $h_{1,2,5}=-\xi_{1,2,5}$, we find only two variables $k_y$ and
$k_z$. Thus, one of $\xi_{1,2,5}$ should depend on the other two. By solving the equations
$h_{1,2,5}=-\xi_{1,2,5}$, we find  gapless points at
$$
\mathbf{k}a=\pm(0,\frac{1}{\sqrt[4]{3}}\sqrt{\frac{\xi_2}{\beta_c \gamma_2}},-\frac{1}{\sqrt[4]{3}}\sqrt{\frac{\gamma_2\beta_c}{ \xi_2}}\frac{\xi_5}{2\gamma_3\beta_c} )
$$
for $\xi_3=\xi_4=0,\xi_5\neq 0, \xi_2 \gamma_2 \beta_c > 0, \xi_1=\frac{1}{\sqrt{3}}(\xi_2-\frac{\gamma_2^2\xi_5^2}{2\gamma_3^2\xi_2})$.

Similarly, if $\xi_3=\xi_5=0,\xi_4\neq 0$, the gapless points are at
$$
\mathbf{k}a=\pm(\frac{1}{\sqrt[4]{3}}\sqrt{-\frac{\xi_2}{\beta_c \gamma_2}},0,-\frac{1}{\sqrt[4]{3}}\sqrt{-\frac{\beta_c\gamma_2}{ \xi_2}}\frac{\xi_4}{2\gamma_3\beta_c} )
$$
for $\xi_2 \gamma_2 \beta_c < 0,  \xi_1=-\frac{1}{\sqrt{3}}(\xi_2-\frac{\gamma_2^2\xi_4^2}{2\gamma_3^2\xi_2})$.

And if $\xi_4=\xi_5=0,\xi_3\neq 0$, the gapless points are at
$$
\mathbf{k}a=\pm(-\frac{\xi_3}{2\sqrt{3}\beta_c\gamma_3}\sqrt{\frac{2\sqrt{3}\beta_c\gamma_2}{\sqrt{3}\xi_1+\xi_2}},\sqrt{\frac{\sqrt{3}\xi_1+\xi_2}{2\sqrt{3}\beta_c\gamma_2}},0)
$$
for $(\sqrt{3}\xi_1-\xi_2)\beta_c\gamma_2>0,(\sqrt{3}\xi_1+\xi_2)\beta_c\gamma_2>0,\gamma_2^2 \xi_3^2+\gamma_3^2(\xi_2^2-3\xi_1^2)=0$.

{\bf Case III}: For $\xi_3$, $\xi_4$ and $\xi_5$, one of them is zero and the other two are non-zero.

Gapless points cannot exist for this case. This is because once two of $\xi_{3,4,5}$
are nonzero, none of three components $k_{x,y,z}$ of a gapless point can be zero. Thus, the other $\xi$
should also be non-zero for gapless points to exist.

{\bf Case IV}: All three $\xi_{3,4,5}$ are non-zero.

For this case, we need to solve five equations $h_i=-\xi_i$ ($i=1,\dots,5$) with three
variables $k_{x,y,z}$. Therefore, only three of the five $\xi$'s are independent.
Let's assume $\xi_{3,4,5}$ are independent variables and one can first solve three equations
$h_{3,4,5}=-\xi_{3,4,5}$ for the momentum $(k_x,k_y,k_z)$ to get possible positions of gapless points
, and then plug them in $\xi_{1,2}=-h_{1,2}$ to get the relation between $\xi_1$ and $\xi_2$  and the exact possible positions.
Results are listed below.

If $\xi_3\neq 0,\xi_4\neq 0,\xi_5\neq 0, \xi_3\xi_4\xi_5\beta_c\gamma_3<0, \xi_3\beta_c\gamma_3<0, \xi_4\beta_c\gamma_3<0, \xi_5\beta_c\gamma_3<0,\xi_2=\frac{\gamma_2 \text{$\xi $3} (\text{$\xi $4}-\text{$\xi $5}) (\text{$\xi $4}+\text{$\xi $5})}{2 \gamma_3 \text{$\xi $4} \text{$\xi $5}}, \xi_1=-\frac{\gamma_2 \left(\text{$\xi $3}^2 \text{$\xi $4}^2+\text{$\xi $3}^2 \text{$\xi $5}^2-2 \text{$\xi $4}^2 \text{$\xi $5}^2\right)}{2 \sqrt{3} \gamma_3 \text{$\xi $3} \text{$\xi $4} \text{$\xi $5}}$,  the gapless points are located at
$$
\mathbf{k}a=\pm\frac{1}{\sqrt{2} \sqrt[4]{3}}( \sqrt{-\frac{\xi_3\xi_4}{\beta_c\gamma_3\xi_5}}, \sqrt{-\frac{\xi_3\xi_5}{\beta_c\gamma_3\xi_4}},\sqrt{-\frac{\xi_4\xi_5}{\beta_c\gamma_3\xi_3}})
$$.

If $\xi_3\neq 0,\xi_4\neq 0,\xi_5\neq 0, \xi_3\xi_4\xi_5\beta_c\gamma_3<0, \xi_3\beta_c\gamma_3<0, \xi_4\beta_c\gamma_3>0, \xi_5\beta_c\gamma_3>0,\xi_2=\frac{\gamma_2 \text{$\xi $3} (\text{$\xi $4}-\text{$\xi $5}) (\text{$\xi $4}+\text{$\xi $5})}{2 \gamma_3 \text{$\xi $4} \text{$\xi $5}}, \xi_1=-\frac{\gamma_2 \left(\text{$\xi $3}^2 \text{$\xi $4}^2+\text{$\xi $3}^2 \text{$\xi $5}^2-2 \text{$\xi $4}^2 \text{$\xi $5}^2\right)}{2 \sqrt{3} \gamma_3 \text{$\xi $3} \text{$\xi $4} \text{$\xi $5}}$,  the gapless points are
$$
\mathbf{k}a=\pm\frac{1}{\sqrt{2} \sqrt[4]{3}}(\sqrt{-\frac{\xi_3\xi_4}{\beta_c\gamma_3\xi_5}},\sqrt{-\frac{\xi_3\xi_5}{\beta_c\gamma_3\xi_4}}, -\sqrt{-\frac{\xi_4\xi_5}{\beta_c\gamma_3\xi_3}})
$$.

If $\xi_3\neq 0,\xi_4\neq 0,\xi_5\neq 0, \xi_3\xi_4\xi_5\beta_c\gamma_3<0, \xi_3\beta_c\gamma_3>0, \xi_4\beta_c\gamma_3<0, \xi_5\beta_c\gamma_3>0,\xi_2=\frac{\gamma_2 \text{$\xi $3} (\text{$\xi $4}-\text{$\xi $5}) (\text{$\xi $4}+\text{$\xi $5})}{2 \gamma_3 \text{$\xi $4} \text{$\xi $5}}, \xi_1=-\frac{\gamma_2 \left(\text{$\xi $3}^2 \text{$\xi $4}^2+\text{$\xi $3}^2 \text{$\xi $5}^2-2 \text{$\xi $4}^2 \text{$\xi $5}^2\right)}{2 \sqrt{3} \gamma_3 \text{$\xi $3} \text{$\xi $4} \text{$\xi $5}}$,  the gapless points are
$$
\mathbf{k}a=\pm\frac{1}{\sqrt{2} \sqrt[4]{3}}(\sqrt{-\frac{\xi_3\xi_4}{\beta_c\gamma_3\xi_5}}, -\sqrt{-\frac{\xi_3\xi_5}{\beta_c\gamma_3\xi_4}},\sqrt{-\frac{\xi_4\xi_5}{\beta_c\gamma_3\xi_3}})
$$.

If $\xi_3\neq 0,\xi_4\neq 0,\xi_5\neq 0, \xi_3\xi_4\xi_5\beta_c\gamma_3<0, \xi_3\beta_c\gamma_3>0, \xi_4\beta_c\gamma_3>0, \xi_5\beta_c\gamma_3<0,\xi_2=\frac{\gamma_2 \text{$\xi $3} (\text{$\xi $4}-\text{$\xi $5}) (\text{$\xi $4}+\text{$\xi $5})}{2 \gamma_3 \text{$\xi $4} \text{$\xi $5}}, \xi_1=-\frac{\gamma_2 \left(\text{$\xi $3}^2 \text{$\xi $4}^2+\text{$\xi $3}^2 \text{$\xi $5}^2-2 \text{$\xi $4}^2 \text{$\xi $5}^2\right)}{2 \sqrt{3} \gamma_3 \text{$\xi $3} \text{$\xi $4} \text{$\xi $5}}$,  the gapless points are
$$
\mathbf{k}a=\pm\frac{1}{\sqrt{2} \sqrt[4]{3}}(-\sqrt{-\frac{\xi_3\xi_4}{\beta_c\gamma_3\xi_5}},\sqrt{-\frac{\xi_3\xi_5}{\beta_c\gamma_3\xi_4}},\sqrt{-\frac{\xi_4\xi_5}{\beta_c\gamma_3\xi_3}})
$$.

\section{Dirac Point With Mirror Symmetry}
\label{app:verification_DP}
In this section, we will consider the low energy effective theory around the gapless points
of the Hamiltonian $H_0+H_{AFM}$ and show it can be described by a Dirac Hamiltonian for the parameter choices discussed
in the Sec.\ref{sec:Dirac Semimetal Phase and Topological Mirror Insulating Phase} in the main text.
Here we consider a generic gapless point $(K_p,K_p,K_z)$ in the momentum space, which exist on a mirror or glide plane.
We expand the momentum around these gapless points with
$(k_x,k_y,k_z)\equiv (K_p+q_x,K_p+q_y,K_z+q_z)$ where $q_{x,y,z}$ are assumed to take small numbers for
perturbation. By expanding $H_D\equiv H_0+H_{AFM}$ around the gapless points $(K_p,K_p,K_z)$ to the first order of $q_i$'s,
we obtain
$$
H_D(K_p+q_x,K_p+q_y,K_z+q_z)\approx\Gamma_0+q_x \Gamma_x+q_y \Gamma_y+q_z \Gamma_z,
$$
where
$\Gamma_0=\frac{4}{15}[E_v+\xi_0-2 \beta_c \gamma_1 K_p^2-\beta_c \gamma_1 K_z^2](J_x^2+J_y^2+J_z^2)$ ,

$\Gamma_x\equiv -\frac{2}{3} \beta_c \gamma_2 K_p(2J_z^2-J_x^2-J_y^2)+2 \beta_c \gamma_2  K_p (J_x^2-J_y^2)+4 \beta_c \gamma_3 K_p J_{xy}+4 \beta_c \gamma_3 K_z J_{zx}$,

$\Gamma_y\equiv -\frac{2}{3} \beta_c \gamma_2 K_p(2J_z^2-J_x^2-J_y^2)-2 \beta_c \gamma_2  K_p (J_x^2-J_y^2)+4 \beta_c \gamma_3 K_p J_{xy}+4 \beta_c \gamma_3 K_z J_{yz}$, and

$\Gamma_z\equiv +\frac{4}{3} \beta_c \gamma_2 K_z(2J_z^2-J_x^2-J_y^2)+4 \beta_c \gamma_3 K_p J_{zx}+4 \beta_c \gamma_3 K_p J_{yz}$.

Since $\beta_c,\gamma_2,\gamma_3 \neq 0$ and $K_p,K_z$ cannot be zero at the same time, none of $\Gamma_x$, $\Gamma_y$ and $\Gamma_z$ are vanishing.
Let us define $q_1\equiv q_x+a_1 q_y +a_2 q_z$, $q_2\equiv q_y +a_3 q_z$, $q_3\equiv q_z$, $\Gamma_1\equiv\Gamma_x$, $\Gamma_2\equiv -a_1\Gamma_x+\Gamma_y$, $\Gamma_3\equiv -a_3\Gamma_x-a_3\Gamma_y+\Gamma_z$
with $a_1\equiv\frac{K_p^2 \left(3 \gamma_3^2-2 \gamma_2^2\right)}{4 \gamma_2^2 K_p^2+3 \gamma_3^2 \left(K_p^2+K_z^2\right)}$, $a_3\equiv\frac{K_p K_z \left(-2 \gamma_2^2+3 \gamma_3^2\right)}{2 K_p^2 \left(\gamma_2^2+3 \gamma_3^2\right)+3 \gamma_3^2 K_z^2}$ and $a_2\equiv a_3(1+a_1)$.
We find that the Hamiltonian $H_D$ can be re-written as
$$
H_D(K_p+q_x,K_p+q_y,K_z+q_z)\approx\Gamma_0+q_1 \Gamma_1+q_2 \Gamma_2+q_3 \Gamma_3,
$$
where $\{\Gamma_i,\Gamma_j\}=0,i\neq j$ and $\{\Gamma_i,\Gamma_i\}\neq 0$ for $i,j=1,2,3$.
Thus, we conclude that all gapless points, if exist, are Dirac points with the low energy effective
theory described by Dirac fermions, for the parameter choices discussed
in the Sec.\ref{sec:Dirac Semimetal Phase and Topological Mirror Insulating Phase} in the main text.

\section{$C_{3v}^*$ spin double group }
\label{app:c3v_double_group}
\begin{table}[h!]
\centering
\begin{tabular}{|c|c|c|c|c|}
 \hline
 $C_{3v}^*$& E & R & $C_3$ & $\sigma_v$ \\
 \hline
 $\Lambda_4$ & 1 & -1 & -1 & i\\
  \hline
 $\Lambda_5$ & 1 & -1 & -1 & -i\\
  \hline
 $\Lambda_6$ & 2 & -2 & 1 & 0\\
 \hline
\end{tabular}
\caption{\label{tab:C3v*}
Character Table of Spin Double Group $C_{3v}^*$\cite{burns2014introduction}}
\end{table}
This section shows how four bases of $\Gamma_8$ bands can be constructed into $\Lambda_{4,5,6}$ irreducible representations of $C_{3v}^*$ spin double group, as discussed in Sec.\ref{sec:TP}. The character table of $C_{3v}^*$ spin double group is shown in Tab.\ref{tab:C3v*}.
Wave functions for each irreducible representations in bases $|\Gamma_8,3/2\rangle$, $|\Gamma_8,1/2\rangle$, $|\Gamma_8,-1/2\rangle$ and $|\Gamma_8,-3/2\rangle$ are shown below:

$\Lambda_4$:
$$N_4\left(\frac{1-i}{\sqrt{2}},\frac{\sqrt{2}-i}{\sqrt{3}},i \left(\frac{1}{\sqrt{6}}+\frac{1}{\sqrt{3}}\right)-\sqrt{\frac{1}{6} \left(3-2 \sqrt{2}\right)},1\right)$$

$\Lambda_5$:
$$
N_5\left(\frac{-1+i}{\sqrt{2}},-\frac{\sqrt{2}+i}{\sqrt{3}},i \sqrt{\frac{1}{6} \left(3-2 \sqrt{2}\right)}-\frac{1}{\sqrt{3}}-\frac{1}{\sqrt{6}},1\right)
$$

$\Lambda_6$:
$$
\frac{1}{\sqrt{6}}\left(-1-i,i \sqrt{3},0,1\right),
\frac{1}{\sqrt{6}}\left(i \sqrt{3},1-i,1,0\right)
$$
,where $N_4$ and $N_5$ are normalization factors.
\section{Tables for Parameters}
\label{app:parameters}
This section is devoted to a summary of all the parameters for the four-band model
and the extended Kane model used throughout the main text of the manuscript.

\begin{table}[H]
$$
\begin{array}{|c|c|c|c|c|c|c|}
\hline
\frac{E_v+\xi_0}{\varepsilon_0}& \gamma_1& \gamma_2&\gamma_3&\frac{C}{\beta_c/a}&\frac{\xi_2}{\varepsilon_0}&\frac{\xi_1}{\varepsilon_0}\\
\hline
0&0&0.5&0.1&0&0&-0.16\\
\hline
\end{array}
$$
\caption{
\label{tab:Diracfig}
Choices of parameters for Fig.\ref{Dirac}a. $\xi_4=\xi_5$,
$\xi_3/\varepsilon_0\in [-0.3,0.3]$ and $\xi_4/\varepsilon_0\in [-0.3,0.3]$,
where $a$ is a real parameter with the unit of length.}
\end{table}

\begin{table}[H]
$$
\begin{array}{|c|c|c|c|c|c|c|}
\hline
\frac{E_v+\xi_0}{\varepsilon_0}& \gamma_1& \gamma_2&\gamma_3&\frac{C}{\beta_c/a}&\frac{\xi_2}{\varepsilon_0}&\frac{\xi_1}{\varepsilon_0}\\
\hline
0&0&0.5&0.1&0.2&0&-0.16\\
\hline
\end{array}
$$
\caption{
\label{tab:Weylfig}
Choices of parameters for Fig.\ref{WeylPD}a. $\xi_4=\xi_5$,
$\xi_3/\varepsilon_0\in [-0.3,0.3]$ and $\xi_4/\varepsilon_0\in [-0.3,0.3]$.}
\end{table}

\begin{table}[H]
$$
\begin{array}{|c|c|c|c|c|c|c|c|c|}
\hline
\frac{E_v+\xi_0}{\varepsilon_0}& \gamma_1& \gamma_2&\gamma_3&\frac{C}{\beta_c/a}&\frac{\xi_2}{\varepsilon_0}&\frac{\xi_3}{\varepsilon_0}&\frac{\xi_4}{\varepsilon_0}&\frac{\xi_1}{\varepsilon_0}\\
\hline
0&0&1&0.5&0.2&0&0&0.1&0.7\\
\hline
\end{array}
$$
\caption{
\label{tab:Nodalfig}
Choices of parameters for Fig.\ref{Nodal}. $\xi_4=\xi_5$.}
\end{table}

\begin{table}[H]
$$
\begin{array}{|c|c|c|c|c|c|c|c|}
\hline
\frac{E_v+\xi_0}{\varepsilon_0}& \gamma_1& \gamma_2&\gamma_3&\frac{C}{\beta_c/a}&\frac{\xi_2}{\varepsilon_0}&\frac{\xi_1}{\varepsilon_0}&\frac{\xi}{\varepsilon_0}\\
\hline
0&0&0.5&0.1&0.2&0&0&-0.025\\
\hline
\end{array}
$$
\caption{
\label{tab:TPfig}
Choices of parameters for Fig.\ref{TPSP}.
$\xi_3=\xi_4=\xi_5\equiv \xi$.}
\end{table}

\begin{table}[H]
$$
\begin{array}{|c|c|c|c|c|c|c|}
\hline
\frac{E_c}{\varepsilon_0} & \frac{E_v+\xi_0}{\varepsilon_0}&\gamma_1&\gamma_2&\gamma_3&\frac{C}{\beta_c/a} & \frac{P}{\beta_c/a}\\
\hline
-2 & 0 & 2+\delta & 0.5+\frac{\delta}{2} & 0.1+\frac{\delta}{2} & 0.2 & 1 \\
\hline
\frac{\xi_3}{\varepsilon_0}&\frac{\xi_4}{\varepsilon_0}&\frac{\xi_3}{\varepsilon_0}& B^+_{8v}/\beta_c &B^-_{8v}/\beta_c & & \\
\hline
 0&0 & 0 & 0 & 0 & & \\
\hline
\end{array}
$$
\caption{
\label{tab:6BPD0(a)}
Choices of parameters for Fig.\ref{6BPD0}a. Here
$\delta\equiv\frac{P^2}{3(E_v+\xi_0-E_c)\beta_c}$, $\xi_1/\varepsilon_0\in (-0.9,0.9)$, $\xi_2/\varepsilon_0\in (-0.9,0.9)$.}
\end{table}

\begin{table}[H]
$$
\begin{array}{|c|c|c|c|c|c|c|}
\hline
\frac{E_c}{\varepsilon_0} & \frac{E_v+\xi_0}{\varepsilon_0}&\gamma_1&\gamma_2&\gamma_3&\frac{C}{\beta_c/a} & \frac{P}{\beta_c/a}\\
\hline
-2 & 0 & 2+\delta & 0.5+\frac{\delta}{2} & 0.1+\frac{\delta}{2} & 0.2 & 1 \\
\hline
\frac{\xi_2}{\varepsilon_0}&\frac{\xi_4}{\varepsilon_0}&\frac{\xi_3}{\varepsilon_0}& B^+_{8v}/\beta_c &B^-_{8v}/\beta_c & & \\
\hline
 0&0 & 0 & 0 & 0 & & \\
\hline
\end{array}
$$
\caption{
\label{tab:6BPD0(b)}
Choices of parameters for Fig.\ref{6BPD0}b. Here
$\delta\equiv\frac{P^2}{3(E_v+\xi_0-E_c)\beta_c}$, $\xi_1/\varepsilon_0\in (-0.9,0.9)$, $\xi_3/\varepsilon_0\in (-0.9,0.9)$.}
\end{table}

\begin{table}[H]
$$
\begin{array}{|c|c|c|c|c|c|c|}
\hline
\frac{E_c}{\varepsilon_0} & \frac{E_v+\xi_0}{\varepsilon_0}&\gamma_1&\gamma_2&\gamma_3&\frac{C}{\beta_c/a} & \frac{P}{\beta_c/a}\\
\hline
-2 & 0 & 2+\delta & 0.5+\frac{\delta}{2} & 0.1+\frac{\delta}{2} & 0.2 & 1 \\
\hline
\frac{\xi_3}{\varepsilon_0}&\frac{\xi_4}{\varepsilon_0}&\frac{\xi_3}{\varepsilon_0}& B^+_{8v}/\beta_c &B^-_{8v}/\beta_c & & \\
\hline
 -0.01& 0.023 & 0.026 & 0 & 0 & & \\
\hline
\end{array}
$$
\caption{
\label{tab:6BPDF}
Choices of parameters for Fig.\ref{6BPDF}a.Here
$\delta\equiv\frac{P^2}{3(E_v+\xi_0-E_c)\beta_c}$, $\xi_1/\varepsilon_0\in (-0.9,0.9)$, $\xi_2/\varepsilon_0\in (-0.9,0.9)$ and the choice of $(\xi_3,\xi_4,\xi_5)$ breaks the symmetry group $C_{3v}$ and its combination with $T_{\mathbf{a}'_3}$.}
\end{table}

\section{Determination of Critical Lines in Fig.\ref{WeylPD}a}
\label{app:WSM_critical_line}
The section describes how to determine critical lines in Fig.\ref{WeylPD}a in the main text.

First, we hope to demonstrate that the Weyl points are from the $\Pi$ pairs on the $(1\bar{1}0)$ plane
in our case.
The creation or annihilation of Weyl points requires two Weyl points with opposite Chern numbers,
thus only occuring between a $\Pi$ pair(like along path $\gamma$) or a $\Pi S$ pair for the case described in Sec.III(b).
If a $\Pi S$ pair merges, it can only happen on the $(1,-1,0)$ axis
where the Hamiltonian is invariant under $\pi_{1\bar{1}0}S$ and each band is doubly degenerate.
If Hamiltonian is gapless at some point on the $(1,-1,0)$ axis, say $(k',-k',0)$,
it requires the Hamiltonian to be an identity at that point, and implies that $\xi_4=0$, $C k'=0$, $\xi_3=2\sqrt{3} \beta_c \gamma_3 k'^2$ and $\xi_1=2\beta_c \gamma_2 k'^2$.
Since $C\neq 0$ and $\sum_{i=1,,,5}\xi_i^2\neq 0$, the Hamiltonian cannot be gapless on the $(1,-1,0)$ axis,
and therefore Weyl points cannot exist on the $(1,-1,0)$ axis.
We conclude Weyl points can only originate from the $\Pi$ pairs on the $(1\bar{1}0)$ plane for the parameter regions
that we are interested in.

When Weyl points are away from the mirror $(1\bar{1}0)$ plane,
energy bands are gapped in the $(1\bar{1}0)$ plane, where the mirror Chern number
is well-defined and can be computed to be 1.
Thus, the mirror Chern number provides additional characterization of topological property in our system.
The mirror Chern number always changed by 1 between the transition from the phase I to II or from II to III
in Fig. 3(a). This suggests that band gap closing at the transition lines
should occur between two bands in one mirror (or glide) subspace. This allows us to determine
the phase transition lines (blue lines in Fig. 3(a)) by solving gapless condition of a mirror (or glide) subspace:
$$
48 \beta_c \gamma_3^2 \xi_3^2 \xi_1+\left(\xi_4^2-\xi_3^2\right)\left(C\pm\sqrt{C^2-16\sqrt{3}\beta_c\gamma_c\xi_3}\right)^2=0.
$$

\end{document}